\documentclass[preprint,aps,amsmath,amssymb]{revtex4}

\newcommand {\bc}{\begin{center}}
\newcommand {\ec}{\end{center}}
\newcommand {\bea}{\begin{eqnarray}}
\newcommand {\eea}{\end{eqnarray}}
\newcommand {\be}{\begin{equation}}
\newcommand {\ee}{\end{equation}}

\def\lsim{\mathrel{\rlap{\lower4pt\hbox{\hskip1pt$\sim$}}
    \raise1pt\hbox{$<$}}}               
\def\gsim{\mathrel{\rlap{\lower4pt\hbox{\hskip1pt$\sim$}}
    \raise1pt\hbox{$>$}}}

\usepackage[mathscr]{euscript}
\usepackage{graphicx}
\usepackage{xcolor}

\begin{document}


\title{Thermal relaxation and the complete set of second 
order transport coefficients for the unitary Fermi gas
from kinetic theory}

\author{Christian Hall and Thomas Sch\"afer}

\affiliation{Department of Physics, North Carolina State University,
Raleigh, NC 27695}

\begin{abstract}
We compute the complete set of second order transport coefficients 
of the unitary Fermi gas, a dilute gas of spin 1/2 particles 
interacting via an $s$-wave interaction tuned to infinite scattering
length. The calculation is based on kinetic theory and the Chapman-Enskog
method at second order in the Knudsen expansion. We take into 
account the exact two-body collision integral. We extend previous 
results on second order coefficients related to shear stress by 
including terms related to heat flow and gradients of the 
fugacity. We confirm that the thermal relaxation time is given by 
the simple estimate $\tau_\kappa = \kappa m/(c_PT)$ even if the 
full collision kernel is taken into account. Here, $\kappa$ is 
the thermal conductivity, $m$ is the mass of the particles, $c_P$ 
is the specific heat at constant pressure, and $T$ is the temperature. 
\end{abstract}

\maketitle

\section{Introduction}
\label{sec_intro}

 The unitary Fermi gas, a dilute gas of spin 1/2 fermions interacting 
via an $s$-wave interaction tuned to infinite scattering length, is  
a very clean system in which to study transport properties of strongly
correlated quantum matter \cite{Schafer:2009dj,Adams:2012th,
Schaefer:2014awa,Zwerger:2016xma}. These studies shed light on 
the behavior of a diverse set of physical systems, including 
neutron matter \cite{Schmitt:2017efp}, the quark gluon plasma 
\cite{Schafer:2009dj}, and strange metals \cite{Hartnoll:2021ydi}. 
A distinct advantage of the unitary Fermi gas is the possibility 
to perform detailed comparisons between theory and experiment in 
the limit of high temperature, where the effective interaction is 
weak. Based on these studies one can then investigate how transport
properties change as the temperature is lowered and the gas enters 
the strongly correlated regime. 

 The quantity that has been studied most extensively is the 
shear viscosity $\eta$ of the unitary Fermi gas. Experimental 
investigations can be found in \cite{oHara:2002,Kinast:2004b,
Bartenstein:2004,Cao:2010wa,Elliott:2013,Elliott:2013b,Joseph:2015},
theoretical studies have been published in \cite{Massignan:2004,
Bruun:2005,Bruun:2006,Bruun:2007,Rupak:2007vp,Son:2008ye,Herzog:2008wg,
Enss:2010qh,Braby:2010tk,Hofmann:2011qs,Enss:2012,Wlazlowski:2012jb,
Hofmann:2019jcj}, and phenomenological analyses are described in
\cite{Schafer:2007pr,Bluhm:2015bzi,Bluhm:2017rnf}. An analysis of 
hydrodynamic flow in the high temperature regime of the unitary 
Fermi gas gives a viscosity $\eta=(0.265\pm 0.020)(mT)^{3/2}$
\cite{Bluhm:2017rnf}, which agrees very well with the theoretical 
prediction $\eta=15/(32\sqrt{\pi})(mT)^{3/2}\simeq 0.264\, 
(mT)^{3/2}$, see \cite{Bruun:2005} and eq.~(\ref{eta_kin}) 
in the present work. Here, $m$ is the mass of the atoms, and $T$ 
is the temperature of the gas. 

  The high temperature result is reliable for $T\gsim T_F$, where
$T_F=(3\pi^2n)^{2/3}/(2m)$ is the Fermi  temperature of the gas, 
and $n$ is the density. For $T\lsim T_F$ the gas is strongly 
correlated, and the superfluid transition takes place at
$T_c=0.167(13)\,T_F$ \cite{Ku:2011}. At $T=T_c$ the shear 
viscosity to entropy density ratio is $\eta/s = 0.50\pm 0.10$
\cite{Bluhm:2017rnf}, several times larger than the proposed 
universal bound $\eta/s=1/(4\pi)$ \cite{Kovtun:2004de}. The 
experimental situation for $T<T_c$ remains unclear. Experiments 
based on hydrodynamic flow indicate that shear viscosity drops 
sharply as $T$ decreases \cite{Joseph:2015,Hou:2021xra}, but the 
sound attenuation constant, which is a linear combination of shear 
viscosity and thermal conductivity, appears to be roughly constant 
below  $T_c$ \cite{Patel:2019udb}.

  Experiments have also studied other transport coefficients, 
including the spin diffusion constant, thermal conductivity,
and the diffusivity of first and second sound \cite{Patel:2019udb,
Sommer:2011,Baird:2019,Wang:2021,Yan:2022qmm,Li:2024}. Indeed,
experiments have reached a level of accuracy and resolution where
it is possible to probe higher order transport coefficients, see
\cite{Schafer:2009dj,Adams:2012th,Schaefer:2014awa} and 
\cite{Burnett:1935,Garcia:2008,York:2008rr,Baier:2007ix,Chao:2011cy,
Schaefer:2014xma}. Higher order transport coefficients correspond to 
higher order gradient corrections to the conserved currents, which 
become important if the system is probed at shorter time and length
scales. Second order transport coefficients are routinely included 
in simulations of the quark gluon plasma, but they are difficult to
measure in that context. 

  In the present work we compute the complete set of second order 
transport coefficients of the unitary Fermi gas in kinetic theory.
The calculation is based on kinetic theory, and it is reliable for
$T\gsim T_F$. We have previously computed second order transport
coefficients related to shear stresses in the fluid 
\cite{Schaefer:2014xma}. In this work we include the effects of 
heat flow and gradients of the fugacity. The simplest transport
coefficient of this type is the thermal relaxation time 
\cite{Frank:2020}, which determines the rate at which the 
energy current relaxes to Fourier's law of heat conduction. 

 We introduce fluid dynamics and the gradient expansion in 
Sect.~\ref{sec_grad}, and kinetic theory in Sect.~\ref{sec:ce}.
The detailed calculation described in 
Sect.~\ref{sec:first}-\ref{sec:res} is outlined in the final
paragraph of Sect.~\ref{sec:ce}. We provide a brief summary 
and outlook in Sect.~\ref{sec:final}, and collect some useful
formulas in the Appendices \ref{sec:units}-\ref{sec:signs}.

\section{Scales and small parameters}
\label{sec_scales}

  The density of the gas determines a momentum scale, the Fermi 
momentum $k_F$, through the relation $n=k_F^3/(3\pi^2)$. Note 
that this scale can be defined at any temperature, and for 
any interaction strength. The density also defines a temperature 
scale, $k_B T_F=k_F^2/(2m)$. In the following we will set the 
Boltzmann constant $k_B$ and Planck's constant $\hbar$ equal to 
one. A dilute atomic gas is characterized by the condition $k_Fr\ll 1$,
where $r$ is the range of the interaction. The limit $k_Fa\to\infty$, 
where $a$ is the $s$-wave scattering length, is called the unitary 
limit. In this limit the gas is dilute, but very strongly correlated. 

 In the regime $T<T_F$ the gas is a strongly correlated quantum 
fluid, and computing equilibrium or non-equilibrium properties 
requires non-perturbative approaches. Quantum Monte Carlo (QMC) 
methods are successful in predicting thermodynamic properties 
\cite{VanHoucke:2011ux}, but the calculations of non-equilibrium 
quantities is more challenging  \cite{Wlazlowski:2012jb}. The 
unitary gas has a second order phase transition that separates
the normal phase from the superfluid phase. Experiments on trapped 
atomic gases give $T_c/T_F = 0.167(13)$ \cite{Ku:2011}.

 If the temperature is high, $T>T_F$, then the thermal de Broglie 
wave length $\lambda_{\it dB}=[2\pi/(mT)]^{1/2}$ of the atoms is 
short, and the quantum diluteness of the gas $n\lambda_{\it dB}^3$ 
can be used as an expansion parameter. In the case of thermodynamic
properties, this corresponds to the virial expansion. We also 
observe that the fugacity of the gas is $z=e^{\mu/T}\simeq 
(n\lambda_{\it dB}^3)/\nu$, so that the virial expansion is 
equivalent to an expansion in powers of $z$. Here, $\mu$ is 
the chemical potential and $\nu=2$ is the spin degeneracy. 
Experimental results show that the viral expansion describes 
the equation of state for $z\lsim 1$ \cite{Nascimbene:2009}, 
corresponding to $T\gsim 0.5 T_F$.

 In this regime transport properties of the unitary gas can be 
understood in terms of kinetic theory and the Boltzmann equation. 
Kinetic theory is based on the existence of well-defined 
quasi-particles, and requires that the width $\Gamma_p$ of 
a quasi-particle with momentum $p$ is small compared to its 
energy $E_p$. Computing the fermion self-energy in the regime
$T\gg T_F$ we find that the condition of applicability of 
kinetic theory is satisfied, $\Gamma_p\sim zT \ll E_p \sim T$
\cite{Dusling:2013sea}. Indeed, it has been demonstrated that 
the kinetic theory result for the shear viscosity can be derived 
from resummed many-body diagrams in the limit $T\gg T_F$
\cite{Hofmann:2019jcj}.

 We will see below that at leading order in the fugacity 
expansion both the shear viscosity and the thermal conductivity 
are independent of $z$. Higher order corrections are suppressed 
by powers of $z$. Empirical attempts to extract higher order terms
in the fugacity expansion \cite{Bluhm:2017rnf} are consistent with
the hypothesis that the range of convergence of the fugacity
expansion of transport coefficients is similar to that of the
virial expansion, $z\lsim 1$.  

 Experiments on trapped ultra-cold atoms involve a macroscopic number 
of particles, typically on the order of $10^5-10^6$. This implies that
the quasi-particle distribution function varies smoothly over the size
$L$ of the trap. In kinetic theory the microscopic scale is given by 
the mean free path $l_{\it mfp}$, and an expansion parameter that 
controls the macroscopic dynamics is given by the Knudsen number 
${\it Kn}=l_{\it mfp}/L$. The mean free path is $l_{\it mfp}=1/
(n\sigma)$, where $\sigma$ is the two-body cross section. In the 
unitary gas $\sigma=4\pi/k^2$, where $k$ is the momentum transfer. 
In the regime $T\gsim T_F$ the thermal average of the cross section
is $\langle\sigma\rangle=8\pi/(mT)$. The Knudsen number of a gas
of $N$ atoms trapped in a box potential with volume $L^3$ is 
\be
\label{Kn}
{\it Kn} = \frac{3^{2/3}\pi^{1/3}}{16N^{1/3}}
   \, \frac{T}{T_F}\, ,
\ee
which implies that ${\it Kn}\ll 1$ unless the temperature is 
very high. 

 In the regime where ${\it Kn}\ll 1$ kinetic theory is equivalent 
to fluid dynamics. Fluid dynamics is based on the assumption 
that the distribution function is close to local thermodynamic
equilibrium, and that thermodynamic variables vary smoothly over 
the size of the system. Gradients of the thermodynamic variables
generate dissipative corrections to the equations of ideal fluid
dynamics. The expansion parameter which controls the validity of 
fluid dynamics is the ratio of dissipative to ideal contributions 
to the conserved currents. The most important parameter is the 
inverse Reynolds number
\be 
 {\it Re}^{-1}=\frac{\eta}{\rho L u}\, , 
\ee
which measures the ratio of dissipative and ideal terms in 
the stress tensor. Here, $\rho$ is the mass density, and $u$ 
is the fluid velocity. In the regime where kinetic theory is 
valid ${\it Kn}\simeq {\it Re}^{-1}$, and the Knudsen expansion 
in kinetic theory is equivalent to the gradients expansion in 
fluid dynamics. The fluid dynamic description is useful because 
the gradient expansion remains valid even in the regime $T\lsim 
T_F$ where the with of quasi-particle width is comparable to 
the energy, and kinetic theory breaks down.

 The ``unreasonable effectiveness of fluid dynamics'' refers
to the observation that there are a number of low viscosity 
fluids in which the gradient expansion works better than 
expected \cite{Noronha-Hostler:2015wft}. This means that a 
description based on the Navier-Stokes equation is accurate, 
even though first order gradient terms are not small. In order 
to study this issue more carefully it is useful to identify
systems in which second (or higher) order gradient terms
can be measured accurately, and the convergence properties 
of the gradient expansion can be studied. The unitary Fermi 
gas is a good candidate for studies of this type, because
the viscosity is small and the interaction is very simple.

  Consider, as an example, the propagation of sound in a 
unitary Fermi gas. The dispersion relation is 
\cite{Schaefer:2014aia}
\be
 \omega = c_s k - \frac{i}{2}\, \Gamma_s k^2 
   + \frac{1}{8c_s} \left( \frac{16}{3} c_s^2\nu_\pi\tau_\pi
     - \Gamma_s^2\right) k^3 + \ldots\, .
\ee
Here, $c_s^2= (\partial P)/(\partial \rho)_{s/n}$ is the 
speed of sound, where the derivative is taken at constant 
entropy per particle. Also, $\Gamma_s=(4\nu_\pi)/3$ is the 
sound attenuation constant, and $\nu_\pi=\eta/\rho$ is the
momentum diffusivity. We have introduced the shear relaxation
time $\tau_\pi$, which will be defined in Sect.~\ref{sec_grad}.
We have made a number of simplifying assumptions. In order 
to obtain a simple estimate we have neglected the contribution
from heat conduction to $\Gamma_s$, and we have ignored second
order gradient terms that are non-linear in the velocity field,
see eq.~(\ref{del_pi_fin}).

 We can compute the ratio of the leading order gradient 
correction, the $O(k^2)$ term, over the $O(k)$ term that 
originates from ideal fluid dynamics
\be
\label{sound-1}
\frac{O(k^2)}{O(k)} = \frac{\Gamma_sk}{2c_s} = 
 \frac{2\sqrt{2}}{\sqrt{15}} \, 
 \Bigg(\frac{\eta}{n}\Bigg)
 \Bigg(\frac{k}{k_F}\Bigg)
 \Bigg(\frac{T_F}{T}\Bigg)^{1/2}\, .  
\ee
Here, we have used the sound speed in the high temperature 
limit, $c_s^2=(5T)/(3m)$, and we have expressed $k$ and $T$
in units of $k_F$ and $T_F$. We can also estimate the relative
magnitude of second order terms 
\be
\label{sound-2}
\frac{O(k^3)}{O(k)} = 
 \frac{16}{15} \, 
 \Bigg(\frac{\eta}{n}\Bigg)^2
 \Bigg(\frac{k}{k_F}\Bigg)^2
 \Bigg(\frac{T_F}{T}\Bigg)\, ,
\ee
where we have used $\tau_\pi=\eta/P$, see eq.~(\ref{final}).

 In order to get an idea of the magnitude of these parameters
we consider the unitary Fermi gas at $T=0.5\,T_F$, which is at the 
limit of the regime of validity of the fugacity expansion. 
At this temperature we have $\eta/n\simeq 0.6$ \cite{Bluhm:2017rnf}.
The experiment of Patel et al.~measures sound modes with wave
numbers up to $k=0.14\,k_F$ \cite{Patel:2019udb}. For these 
modes the first order ratio in eq.~(\ref{sound-1}) is 0.086, 
and the second order ratio in eq.~(\ref{sound-2}) is 0.015. This
implies that second order effects, which cause a non-linearity
in the sound dispersion relation, are at the limit of what can 
be observed in current experiments. Second order effects can
be enhanced by studying larger wave numbers or higher 
temperatures.

\section{Fluid dynamics and the gradient expansion}
\label{sec_grad}

 Fluid dynamics is based on the conservation laws for mass, momentum,
and energy,
\bea 
\partial_t \rho + \nabla\cdot \vec{\jmath}_m &=& 0\, ,  
\label{hydro-1}\\
\partial_t \pi_i + \nabla_j\Pi_{ij} &=& 0\, ,  
\label{hydro-2}\\
\partial_t {\cal E} + \nabla\cdot \vec{\jmath}_{\epsilon}
   &=& 0 \, .
\label{hydro-3}
\eea
Here, $\rho$ is the density of mass, $\pi_i$ is the density of
momentum, and ${\cal E}$ is the density of energy. The density
of momentum can be used to define the fluid velocity, $\pi_i=\rho
u_i$, and the mass current is fixed by symmetries, $\vec\jmath_m
=\vec\pi$. 

 In order to close the equations we need to determine the 
constitutive equations for the currents $\Pi_{ij}$ and 
$\vec{\jmath}_{\epsilon}$. Fluid dynamics is an effective theory
for the long-time, large-distance dynamics of the system. For 
this reason we can systematically expand the currents in gradients
of the hydrodynamic variables. The stress tensor can be 
written as 
\be 
\label{Pi_0}
 \Pi_{ij} = \rho u_i u_j + P\delta_{ij}+ \delta \Pi_{ij}\, ,
\ee
where $P$ is the pressure and $\delta\Pi_{ij}$ contains higher 
derivative terms. The pressure is fixed by the equation of state,
$P=P({\cal E},\rho)$. In a scale invariant non-relativistic fluid
we have $P=\frac{2}{3}{\cal E}$, and in the high-temperature 
limit $P=nT$ and ${\cal E}=\frac{3}{2}nT$, where $n=\rho/m$ is 
the particle density and $m$ is the mass of the particles. 

 At first order in the gradient expansion $\delta\Pi_{ij}$ is given
by $\delta\Pi_{ij}=-\eta\sigma_{ij}-\zeta\delta_{ij} \langle \sigma
\rangle$ with
\be 
\label{sig_ij}
 \sigma_{ij} = \nabla_i u_j +\nabla_j u_i 
  -\frac{2}{3}\delta_{ij}   \langle\sigma\rangle \, ,
\hspace{0.1\hsize}
 \langle\sigma\rangle =\vec{\nabla}\cdot\vec{u}\, ,
\ee
where $\eta$ is the shear viscosity and $\zeta$ is the bulk viscosity.
We can show that in a scale invariant fluid $\zeta=0$ \cite{Son:2005tj}.
The energy current is given by $\vec\jmath_{\epsilon}=({\cal E}+P)
\vec{u}+\delta\vec{\jmath}_{\epsilon}$. At first order in gradients
\be 
\label{q_i}
 \delta\jmath^i_{\cal \epsilon} =
    u_j\delta\Pi^{ij} -\kappa \nabla^i T\, , 
\ee
where $\kappa$ is the thermal conductivity. We also define $q_i=-
\nabla_i\ln(T)$ so that $\delta\vec{\jmath}_{\epsilon}\sim \kappa 
T\vec{q}$.

 The general form of dissipative terms in a scale invariant 
non-relativistic fluid up to second order in the gradient expansion 
was given in \cite{Chao:2011cy}. The stress tensor is  given by 
\bea 
\delta\Pi_{ij} &=& -\eta\sigma_{ij}
   + \eta\tau_\pi\bigg(
      \dot\sigma_{ij} + u^k\nabla_k \sigma_{ij}
    + \Theta_\pi \langle \sigma\rangle \sigma_{ij} \bigg)
    + \lambda_1 \sigma_{\langle i}^{\;\;\; k}\sigma^{}_{j\rangle k} 
    + \lambda_2 \sigma_{\langle i}^{\;\;\; k}\Omega^{}_{j\rangle k}\nonumber\\
   && \mbox{} 
    + \lambda_3 \Omega_{\langle i}^{\;\;\; k}\Omega^{}_{j\rangle k}  
    + \gamma_1 q_{\langle i}q_{j\rangle}
    + \gamma_2 q_{\langle i}\nabla_{j\rangle}\alpha
    + \gamma_3 \nabla_{\langle i}\alpha\nabla_{j\rangle}\alpha  \nonumber \\[0.1cm]
   \label{del_pi_fin}
   && \mbox{}
    + \gamma_4 \nabla_{\langle i}q_{j\rangle} 
    + \gamma_5 \nabla_{\langle i}\nabla_{j\rangle}\alpha \, . 
\eea
Here, $\tau_\pi$ is a relaxation time, and $\lambda_i$ and $\gamma_i$ 
are eight independent second order transport coefficients. 
${\cal O}_{\langle ij \rangle}=\frac{1}{2}({\cal O}_{ij}+{\cal O}_{ji}
-\frac{2}{3}\delta_{ij}{\cal O}^k_{\;\;k})$ is the symmetric traceless
part of the second rank tensor ${\cal O}_{ij}$, and $\Omega_{ij} 
=\nabla_iu_j-\nabla_ju_i$ denotes the vorticity tensor. The parameter 
$\Theta_\pi$ is determined by conformal symmetry, but the conformal 
constraints on non-relativistic second order fluid dynamics have not 
been carefully studied \footnote{
The analogous constraints are known in the case of relativistic
fluids \cite{Loganayagam:2008is}. In the non-relativistic case
the constraints on first order hydrodynamics are known 
\cite{Son:2005tj,Chao:2011cy}, but the most general form of the 
currents at second order in the gradient expansion has not been 
established, even though the necessary tools have been developed
\cite{Jensen:2014ama}. Note that Chao and Sch\"afer present an 
argument that $\Theta_\pi=2/3$.}.
The energy current is 
\bea
\delta\jmath_{\epsilon}^i &=& u_j\delta\Pi^{ij}+ \kappa T q^i
- \kappa T\tau_\kappa\bigg(
      \dot{q}^{i} + u^k\nabla_k q^i
    + \Theta_\kappa \langle \sigma\rangle q^i \bigg) 
    + \nu_1 \sigma^{ij}q_j
     \nonumber \\
    \label{del_j_fin}
    & & \mbox{}\hspace{1cm}
      + \nu_2 \Omega^{ij}q_j
      + \nu_3 \sigma^{ij}\nabla_j\alpha 
      + \nu_4 \Omega^{ij}\nabla_j\alpha
      + \nu_5 \nabla_j\sigma^{ij}
      + \nu_6 \nabla_j\Omega^{ij}\, , 
\eea
which defines the relaxation time $\tau_\kappa$ as well as six
second order transport coefficients $\nu_i$ and the parameter 
$\Theta_\kappa$. Note that the representation in 
eqs.~(\ref{del_pi_fin},\ref{del_j_fin}) is not unique. First, 
we can choose to represent gradient corrections in terms of 
densities such as $n$ and ${\cal E}$ instead of intensive 
variables such as $\ln(T)$ and $\alpha$. Furthermore, we 
can use the equations of motion to express $\dot\sigma_{ij}$
and $\dot{q}_i$ in terms of spatial gradients. Our choice
here is motivated by the fact that $\alpha$ and $\ln(T)$
are convenient variables in kinetic theory, and that retaining
the time derivatives in $\dot\sigma_{ij}$ and $\dot{q}_i$
makes the physical meaning of $\tau_\pi$ and $\tau_\kappa$
more transparent. 

\section{Kinetic theory and the Knudsen expansion}
\label{sec:ce}

 In the following we will compute the second order transport 
coefficients in the high temperature limit of the the unitary 
gas using kinetic theory. In the high temperature limit the 
theory possesses well defined quasi-particles with energy $E_p
=p^2/(2m)$ and small width $\Gamma_p\ll E_p$ \cite{Dusling:2013sea}.
The conserved currents can be expressed in terms of quasi-particle
distribution functions $f_p(\vec{x},t)$. The currents are 
\bea 
\label{j_m_kin}
\vec{\jmath}_m = \nu\int d\Gamma_p \, m\vec{v}f_p(\vec{x},t)\, , \\
\label{pi_ij_kin}
\Pi_{ij} = \nu \int d\Gamma_p \, v_i p_jf_p(\vec{x},t)\, , \\
\label{j_e_kin}
\vec{\jmath}_{\epsilon} = \nu \int d\Gamma_p \, \vec{v}E_p
    f_p(\vec{x},t)\, .
\eea
Here, $\nu=2$ is the spin degeneracy and $d\Gamma_p=(d^3p)/(2\pi)^3$ is 
the volume element in momentum space. We have also defined the 
quasi-particle velocity $\vec{v}=\vec{\nabla}_pE_p$, where $\vec{\nabla}_p$
is the gradient with respect to $\vec{p}$. The distribution function satisfies 
the Boltzmann equation
\be
\label{be}
\left( \partial_t + \vec{v}\cdot\vec{\nabla}_x 
                  - \vec{F}\cdot\vec{\nabla}_p \right) 
  f_p(\vec{x},t) = C[f_p]\, , 
\ee
where $\vec{F}=-\vec{\nabla}_xE_p$. In the high temperature limit
we compute the transport coefficients at leading order in the fugacity
$z=\exp(\mu/T)$. In this limit $E_p$ is only a function of $\vec{p}$
and not a function of $\vec{x}$ so that $\vec{F}=0$ unless there 
are external forces present. In a medium the energy
$E_p$ is modified by a self energy correction ${\it Re}\,\Sigma(p)
\sim (z/a)F(p)$, where $F(p)$ is a function that has been computed 
in \cite{Dusling:2013sea}. This correction is indeed of higher order
in $z$, and it vanishes at unitarity.

The Boltzmann equation can be written
as ${\cal D}f_p=C[f_p]$, where we have defined the comoving 
derivative in the particle frame
\be 
 {\cal D} = \partial_t + \vec{v}\cdot\vec{\nabla}_x\, . 
\ee
Below, we will also make use of the comoving derivative in the 
fluid frame ${\cal D}_u=\partial_t+\vec{u}\cdot\vec{\nabla}$. 
At leading order in the fugacity $z$ the collision term is dominated 
by two-body collisions. Furthermore, the effects of quantum statistics 
can be neglected. We obtain 
\be 
 C[f_1]= -\prod_{i=2,3,4}\Big(\int d\Gamma_{i}\Big) w(1,2;3,4)
   \left( f_1f_2-f_3f_4\right)\, , 
\ee
where $f_i=f_{p_i}$ and $w(1,2;3,4)$ is the transition probability 
for the elastic two-body process $\vec{p}_1+\vec{p}_2\to\vec{p}_3
+\vec{p}_4$. In the unitary Fermi gas the scattering amplitude is
dominated by $s$-wave scattering and
\be
w(1,2;3,4) = (2\pi)^4\delta\Big(\sum_i E_i\Big)
         \delta\Big(\sum_i \vec{p}_i\Big) \,|{\cal A}|^2\, ,  
 \hspace{0.5cm} 
      |{\cal A}|^2 = \frac{16\pi^2}{m^2}\frac{a^2}{q^2a^2+1}\, . 
\ee
Here, $a$ is the $s$-wave scattering length and $2\vec{q}=\vec{p}_2
-\vec{p}_1$ is the momentum transfer. The collision term vanishes 
in local thermal equilibrium, described by the distribution function
\be 
\label{f_MB}
 f^{0}_p= \exp\Big(\frac{\mu}{T}\Big)
      \exp\Big(-\frac{m\vec{c}^{\,2}}{2T}\Big)\, , 
\ee
where $\vec{c}=\vec{v}-\vec{u}$ is the velocity of the particle 
relative to that of the fluid. In general, the thermodynamic variables
$\mu,T$ and $\vec{u}$ are functions of $\vec{x}$ and $t$.

 In the following we will lay out the formalism for computing transport
coefficients using the method Chapman and Enskog. This method is very 
well established \cite{Chapman:1990}, but analytical calculations
using a microscopic interaction are rare. Calculations at higher order
in the derivative expansion can be simplified by introducing an inner
product on the space of distribution functions, and by making use of 
the fact that the linearized collision operator is a Hermitean operator
on that space. We use the notation adopted in our earlier work 
\cite{Schaefer:2014xma}. For completeness, we repeat these definitions 
below.

 In the Chapman-Enskog procedure we solve the Boltzmann equation by
expanding the distribution function $f_p$ around the local equilibrium
distribution, 
\be 
 f_p= f_p^{0} + f_p^{1} + f_p^{2} + \ldots 
    =  f_p^{0}\Big( 1 + \frac{\psi^{1}_p}{T}  + \frac{\psi^{2}_p}{T} 
     + \ldots \Big) \, . 
\ee
Inserting this expression into the Boltzmann equation gives
\be 
\label{ce}
 {\cal D}f_p^{0} + {\cal D}f_p^{1} + \ldots  
 = \frac{f_p^{0}}{T} \Big( 
      C^{1}_L\left[\psi^{1}_p\right] +  C^{2}_L\left[\psi^{1}_p\right]
    + C^{1}_L\left[\psi^{2}_p\right] + \ldots \Big)\, , 
\ee
where we have expanded the collision term in powers of $\psi^i_p$, 
\bea
  C^{1}_L\left[\psi^i_1\right] &=&  
  -\int\Big(\prod_{i=2,3,4}d\Gamma_{i}\Big) w(1,2;3,4)\, f^0_{p_2}\,
      \left( \psi^i_1+\psi^i_2
         -\psi^i_3-\psi^i_4 \right)\, ,  \\  
  C^{2}_L\left[\psi^i_1\right] &=&
  -\int\Big(\prod_{i=2,3,4}d\Gamma_{i}\Big) w(1,2;3,4)\, 
  \frac{f^0_{p_2}}{T}\,
      \left( \psi^i_1\psi^i_2
         -\psi^i_3\psi^i_4 \right)\, .
\eea
The left hand side of eq.~(\ref{ce}) is an expansion in gradients 
of the thermodynamic variables, and the right hand side contains 
powers of the inverse mean free path $l_{\it mfp}^{-1}$. The
dimensionless expansion parameter is the Knudsen number ${\it Kn}
=l_{\it mfp}/L$. Here, $L$ is a characteristic length scale associated
with the spatial variation of thermodynamic variables. 

 In this work we will provide a complete solution of the Boltzmann 
equation at second order in the Knudsen number. At first order a
formal solution is given by 
\be 
\label{psi_1}
 \psi^{1}_p = \left(C_L^1\right)^{-1} X^0_p\, ,\hspace{1cm}
 X^0_p\equiv \frac{T}{f^0_p}\left({\cal D}f^0_p\right)\, . 
\ee
Once the function $\psi^1_p$ has been determined the dissipative
contribution to the stress tensor at first order in the gradient
expansion can be computed using 
\be 
\label{pi_ij_1}
\delta \Pi^1_{ij} = \frac{\nu m}{T}\langle v_iv_j | \psi^1_p\rangle \, , 
\ee
where we have defined the inner product 
\be
\label{braket}
 \langle \psi_p|\chi_p\rangle = \int d\Gamma_p \, f^0_p \,
          \psi_p\chi_p \, .
\ee
The unitary gas is scale invariant and $\delta \Pi_{ij}$ is traceless. 
This implies that we can replace $v_{ij}\equiv v_iv_j$ by $\bar{v}_{ij}
\equiv v_{ij}-\frac{\delta_{ij}}{3}v^2$. The first order gradient term
in the energy current is given by 
\be
\label{delta-j1}
\delta j^1_{\epsilon, i} 
  = \frac{\nu}{2mT} \langle v_ip^2 | \psi^1_p\rangle \, . 
\ee
At second order in the Knudsen expansion the formal solution of the 
Boltzmann equation can be written as
\be 
\label{psi_2}
\psi^2_p =  \left(C_L^1\right)^{-1} \Big\{
   \left( X^0_p-C^1_L\left[\psi^1_p\right]\right)
 + X^1_p - C_L^2\left[\psi^1_p\right] \Big\}
\, ,\hspace{0.4cm}
 X^1_p\equiv \frac{T}{f^0_p}\left({\cal D}f^1_p\right)\, . 
\ee
This result is difficult to use in practice, because we have to 
determine the action of the inverse collision operator on the 
complicated expression in the curly bracket. The calculation is 
simplified by the observation that the second order contribution
to the conserved currents is given by an inner product of $\psi^2_p$
with $\bar{v}_{ij}$ and $v_i p^2$. We can make use of the fact 
that the symmetries of the collision term imply that $C_L^1$ is a
a Hermitean operator with respect to this inner product. We can 
write 
\be 
\label{pi_ij_2}
 \delta \Pi^2_{ij} = \frac{\nu m}{T}
  \big\langle \left(C_L^1\right)^{-1}\bar{v}_{ij} \big|
  \Delta X^0_p + X^1_p - C_L^2\left[\psi^1_p\right] \big\rangle \, ,
\ee
where $\Delta X^0_p = X^0_p-C^1_L\left[\psi^1_p\right]$. Note that  
eq.~(\ref{psi_1}) implies that $\Delta X^0_p$ vanishes at first order 
in the gradient expansion. However, in general $\Delta X^0_p$ is 
non-vanishing at $O(\nabla^2)$. The advantage of having the inverse
collision operator act to the left is that the action of $(C_L^1)^{-1}$
on $\bar{v}_{ij}$ is determined by the first order equation $\psi^1_p
=(C_L^1)^{-1}X_p^0$. The second order expression for the energy current 
is given by 
\be 
\label{delta-j2}
 \delta j^2_{\epsilon,i} = \frac{\nu}{2mT}
  \big\langle \left(C_L^1\right)^{-1} v_i p^2 \big|
  \Delta X^0_p + X^1_p - C_L^2\left[\psi^1_p\right] \big\rangle \, .
\ee
In the remainder of this work we will compute the various ingredients
in eq.~(\ref{pi_ij_2}) and (\ref{delta-j2}). In Sect.~\ref{sec:first}
we solve the first order equation for $\psi^1_p$. This solution was 
previously obtained for the shear part in \cite{Bruun:2005} and for 
the thermal conduction term in \cite{Braby:2010ec}. In Sect.~\ref{sec:X1}
we compute the second order streaming terms $X^1_p$. A subset of these
terms was previously computed in \cite{Schaefer:2014xma}, and results 
in the relaxation time approximation can be found in \cite{Chao:2011cy}.
In Sect.~\ref{sec:coll} we determine the second order collision term
$C_L^2\left[\psi^1_p\right]$. In Sect.~\ref{sec:delj2-pi2} we compute 
the matrix elements defined in eqs.~(\ref{pi_ij_2},\ref{delta-j2}). 
Finally, in Sect.~\ref{sec:ct} we show that $\Delta X_p^0$ does not
contribute to the transport coefficients. The results for the 
transport coefficients are summarized in Sect.~\ref{sec:res}.

\section{First order solution of the Boltzmann equation}
\label{sec:first}

 Consider the first order streaming term $X^0_p= \frac{T}{f^0_p}
({\cal D}f^0_p)$. Using eq.~(\ref{f_MB}) we get
\bea
\label{X^0}
X^0_p & =& \frac{m}{2}\, \bigg\{ 
    \frac{2T}{m}\, {\cal D}_u \alpha 
 + 2c^i \left[ {\cal D}_u u_i + \frac{T}{m} \nabla_i\alpha \right]  
   \nonumber \\
 & & \hspace{0.7cm}\mbox{} 
 + c^ic^j\left[ \sigma_{ij} + \delta_{ij} \left( {\cal D}_u \ln(T) 
      + \frac{2}{3}\langle \sigma\rangle \right)\right]
 + c^2c^k\nabla_k \ln(T) \bigg\}\, .  
\eea
Here, we have used the identities collected in App.~\ref{sec:formulas}.
We have also introduced the comoving fluid derivative ${\cal D}_u
=\partial_0+\vec{u}\cdot\vec{\nabla}$ mentioned above. Finally, we
we have defined $\alpha=\mu/T$. Eq.~(\ref{X^0}) can be simplified 
using the equations of fluid dynamics. In App.~\ref{sec:NS} we
show how to write these equations in terms of the intensive variables
$\alpha$, $u_i$ and $\ln(T)$. At leading order in the fugacity 
expansion we can use the equation of state of a free gas, $P=nT$. 
At first order in gradients we get 
\be 
\label{euler}
 {\cal D}_u\alpha=0\, , \hspace{0.6cm}
 {\cal D}_uu_i + \frac{T}{m} \nabla_i\alpha = 
   -\frac{5T}{2m}\nabla_i\ln(T)\, , \hspace{0.6cm}
 {\cal D}_u\ln(T) + \frac{2}{3}\langle \sigma\rangle=0\, ,
\ee
which gives
\be
\label{X^0_eq2}
X^0_p = \frac{m}{2}\, \bigg\{ 
   c^ic^j \sigma_{ij} 
  - \frac{c^k}{T} \left[\frac{5T}{m}- c^2 \right] \nabla_k T \bigg\}
  \equiv (X^0_p)^{ij}\sigma_{ij} + (X^0_p)^k\nabla_k T\, . 
\ee
We observe that eq.~(\ref{X^0_eq2}) is orthogonal to the zero modes 
of the collision operator, $\langle X^0_p |\chi^{i}_{\it zm}\rangle$
with $\chi^{i}_{\it zm} =\{1,\vec{c},c^2\}$ for $i=1,2,3$. The zero 
modes are associated with conservation of particle number, momentum, 
and energy. 
 
 In order to solve the Boltzmann equation $|X^0_p\rangle 
=C_L^1 |\psi^1_p\rangle$ we split $\psi^1_p$ into a shear part 
and a term related to heat flow, 
\be
\psi^1_p=(\psi^1_p)^{\it sh} + (\psi^1_p)^{\it th}
 \equiv (\psi^1_p)^{ij}\sigma_{ij} + (\psi^1_p)^k \nabla_k T.
\ee
We write both terms as series expansions in terms of a suitable
set of complete functions. For the shear part we write 
\be 
 (\psi^1_p)^{ij} = \sum_{k=0}^{N-1} a_k S_k\left(x_c\right)
   \bar{c}^{ij}\, , 
   \hspace{0.5cm}
    x_c = \frac{mc^2}{2T}\, . 
\label{psi1_sh}
\ee
In the case of shear we choose $S_k(x)=L_k^{5/2}(x)$, where 
$L_k^{5/2}$ is a generalized Laguerre polynomial of order $5/2$. 
This choice is convenient because of the orthogonality relation 
$\langle S_k\,\bar{c}^{ij}|S_l\, \bar{c}_{ij}\rangle\sim \delta_{kl}$.
Note that $S_0(x)=1$. The coefficients $a_k$ are determined by moments 
of the Boltzmann equation
\be 
\label{BE_mom}
 \big\langle S_k(x_c)\, \bar{c}^{ij} \big| (X^0_p)_{ij} \big\rangle 
  =  \big\langle S_k(x_c)\, \bar{c}^{ij} \big| C_L^1 
         \big| (\psi^1_p)_{ij}\big\rangle \, ,
  \hspace{0.5cm}(k=0,\ldots, N-1) \, .
\ee
Eq.~(\ref{X^0_eq2}) implies that $(X^0_p)^{ij}=\frac{m}{2}\,\bar{c}^{ij}$.
The first order in the Laguerre polynomial expansion corresponds to 
setting $N=1$. Then $(\psi^1_p)^{\it sh}=a_0\,\bar{c}^{ij}\sigma_{ij}$
where $a_0$ is given by
\be 
a_0 =\frac{m}{2}\,\frac{\big\langle \bar{c}^{kl} \big|\bar{c}_{kl}\big\rangle}
 {\big\langle \bar{c}^{ij} \big| C_L^1  \big| \bar{c}_{ij}\big\rangle}\, . 
\label{a0}
\ee
The numerator contains the matrix element of the linearized collision 
operator, given by  
\be 
\big\langle \bar{c}_{ij} \big| C_L^1  \big| \bar{c}^{ij}\big\rangle
 =  -\int\Big(\prod_{i=1}^4 d\Gamma_{i}\Big) w(1,2;3,4)\, f^0_1 f^0_2\,
     \left(\bar{c}_1\right)_{ij}\, 
         \left( \bar{c}_1^{ij}+\bar{c}_2^{ij}-\bar{c}_3^{ij} - \bar{c}_4^{ij}  
         \right)\, ,
\ee
where $mc_k=p_k$. This integral can be computed analytically using the methods 
described in Sect.~\ref{sec:coll}. We get
\be 
\label{a_0_kin}
 a_0 \equiv \bar{a}_0 \, \frac{m}{zT}
     = -\frac{15\pi}{32\sqrt{2}} \, \frac{m}{zT}\, .
\ee
Using eq.~(\ref{pi_ij_1}) we can now compute the first order gradient
correction $\delta \Pi^1_{ij}$. Matching to the expression in fluid dynamics,
$\delta\Pi_{ij}^1=-\eta\sigma_{ij}$ determines the shear viscosity,
\be
\label{eta_kin}
 \eta = \frac{15}{32\sqrt{\pi}}\, (mT)^{3/2}\, . 
\ee
Note that this result corresponds to the leading order in an expansion
of $(\psi^1_p)^{ij}$ in Laguerre polynomials. Higher order corrections
are quite small, on the order of 2\% \cite{Schaefer:2014xma}. Finally,
we note that the Boltzmann equation, $(C_L^1)^{-1}|(X^0_p)_{ij}\rangle
= |(\psi^1_p)_{ij}\rangle$, together with the explicit form of 
$(X^0_p)_{ij}$ and $(\psi^1_p)_{ij}$ determines the action of 
$(C_L^1)^{-1}$ on $|\bar{v}_{ij}\rangle$. We have
\be 
\label{C_L_inv}
\left(C_L^1\right)^{-1}\big|\bar{v}_{ij}\big\rangle = 
\big|\bar{v}_{ij}\big\rangle  \frac{2}{zT}\,\bar{a}_0  \, , 
\ee
which we will use to compute the matrix element in eq.~(\ref{pi_ij_2}).
Note that equ.~(\ref{C_L_inv}) defines a collision time for shear 
relaxation, $\tau_R^{\it sh}=2|\bar{a}_0|/(zT)$. 

 Next, we focus on the heat flow term in $X^0_p$ and solve for the 
thermal conductivity $\kappa$. We expand the thermal part of $\psi^1_p$
as
\be 
 (\psi^1_p)^i = c^i \sum_{k=1}^{N} b_k R_{k}\left(x_c\right)
   \hspace{0.5cm}
    x_c = \frac{m c^2}{2T}\, . 
\label{psi1_th}
\ee
Here we have defined $R_k(x)=L_k^{3/2}(x)$, where $L_k^{3/2}$ 
are generalized Laguerre polynomial of order $3/2$. This choice is
ensures the validity of the orthogonality relation $\langle R_k
\, c^{i}|R_l\, c_{i}\rangle \sim \delta_{kl}$. We exclude $k=0$
in the sum in order to ensure orthogonality to the zero mode 
$\chi_{\it zm}=\vec{c}$. The lowest order term is $R_1(x)=
\frac{5}{2}-x$. As in the shear case the coefficients $b_k$ are 
determined by moments of the Boltzmann equation
\be 
\label{BE_mom_th}
 \big\langle R_k(x_c)\, c^{i} \big| (X^0_p)_{i} \big\rangle 
  =  \big\langle R_k(x_c)\, c^{i} \big| C_L^1 
         \big| (\psi^1_p)_{i}\big\rangle \, ,
  \hspace{0.5cm}(k=1,\ldots, N)\, . 
\ee
At leading order $(N=1)$ we obtain
\be 
b_1 =\frac{\big\langle c_{k} R_{1}(x_c) \big| (X^0_p)^{k}\big\rangle}
          {\big\langle c_{i} R_{1}(x_c) \big| C_L^1 
                         \big| c^iR_1(x_c)\big\rangle}\, . 
\ee
Using $(X_p^0)^k=-R_1 c^k$ we obtain the normalization integral
\be 
\big\langle  c_{k} R_{1}(x_c)  \big| (X^0_p)^{k} \big\rangle 
=-\frac{15 z\sqrt{m}\,T^{5/2}}{4 \sqrt{2}\,\pi^{3/2}}\, . 
\ee
The denominator involves a matrix element of the linearized 
collision operator. We have
\bea
\big\langle  c_{i} R_{1}(x_c)  \big| C_L^1 
         \big| R_1(x_c) c^i\big\rangle
 &=&  -\int\Big(\prod_{i=1}^4 d\Gamma_{i}\Big) 
       w(1,2;3,4)\, f^0_1 f^0_2\, 
        \left(c_1\right)_{i}R_1(x_1)\, 
       \nonumber \\
 &&  \hspace*{1.25cm}
         \left(c_1^{i}R_1(x_1)
         +c_2^{i}R_1(x_2)
         -c_3^{i}R_1(x_3) 
         -c_4^{i}R_1(x_4)  
         \right)\, . 
\eea
The integrand can be simplified using the symmetries of 
the collision term. In particular, the matrix element is 
invariant under the exchange of the initial or final 
particles, $1\leftrightarrow 2$ and $3\leftrightarrow 4$,
as well as under the exchange of the initial and final 
particles, $(12)\leftrightarrow(34)$. We can replace
\bea
&&\left(c_1\right)_{i}R_1(x_1) 
         \left(c_1^{i}R_1(x_1)
         +c_2^{i}R_1(x_2)
         -c_3^{i}R_1(x_3) 
         -c_4^{i}R_1(x_4) \right) 
   \rightarrow \nonumber \\
 && \hspace{4cm}\frac{1}{4} \left(c_1^{i}R_1(x_1)
         +c_2^{i}R_1(x_2)
         -c_3^{i}R_1(x_3) 
         -c_4^{i}R_1(x_4)\right)^2  \, . 
\eea
The collision integral can be evaluated using the methods
described in Sect.~\ref{sec:coll}. We find 
\be
\label{me_CL_th}
\big\langle  c_{i} R_{1}(x_c) \big| C_L^1
                      \big| R_1(x_c) c^i\big\rangle
  = -\frac{8 z^2 \sqrt{m}\ T^{7/2}}{3 \pi ^{5/2}}\, . 
\ee
The minus sign is consistent with the fact that the collision 
operator has negative eigenvalues, and describes a set of 
decay rates. From these results we obtain 
\be
 b_1 \equiv \frac{\bar{b}_1}{zT}
    = \frac{45\pi}{32 \sqrt{2}} \frac{1}{zT}\, .
 \label{A1}
\ee
Matching the energy current to the hydrodynamic result 
$\delta\vec\jmath_{\epsilon}=-\kappa\vec\nabla T$ we obtain
\be 
\kappa= -\frac{\nu b_1}{3T} 
  \big\langle v^{i} E_p  \big| R_1(x_c) c_i\big\rangle 
  = \frac{225}{128 \sqrt{\pi}} \, m^{1/2}T^{3/2} . 
\ee
As in the case of shear, we observe that the Boltzmann equation
determines the action of $(C^1_L)^{-1}$ on $|c^iR_1(x_c)\rangle$
\be 
\label{CL-inv-A1}
 (C^1_L)^{-1} | c^i R_1(x_c)\rangle 
   =-\frac{\bar{b}_1}{zT}  | c^i R_1(x_c)\rangle \, . 
\ee   
This result will be used to evaluate the matrix element in
eq.~(\ref{delta-j2}). We note that eq.~(\ref{CL-inv-A1}) defines 
the collision time for thermal relaxation, $\tau_R^{\it th}=
\bar{b}_1/(zT)$, and that $\bar{b}_1=-2\alpha_0\bar{a}_0$, where 
$\alpha_0=3/2$ is the inverse Prandtl ratio \cite{Braby:2010ec}.

\section{Second order solution: Streaming terms}
\label{sec:X1}

 The second order streaming term is $X_p^1 = (T/f_p^0)(\mathcal{D}
f_p^1)$, where $f_p^1=f_p^0\psi^1_p/T$ and $\psi^1_p$ was determined 
in eq.~(\ref{psi1_sh},\ref{a_0_kin}) for the shear part, and in 
eq.~(\ref{psi1_th},\ref{A1}) for the heat flow. Combining these 
two terms we have 
\be
\psi^1_p = \frac{\bar{a}_0m}{zT}
  \left( \bar{c}^{ij}\sigma_{ij} -\alpha_0 \left( c^2 -\alpha_1\right)
    c^i q_i \right) ,
\label{psi1_all}
\ee
where $\alpha_0=3/2$, $\alpha_1=5T/m$, and $q_i=-\nabla_i\ln(T)$. 
This result can be compared to an earlier calculation based on the 
Boltzmann equation in the relaxation time approximation (RTA). In 
this approximation we take the collision term to be $C[f_p]=-\delta f_p/
\lambda$ \cite{Bhatnagar:1954}. For classical Boltzmann statistics 
eq.~(63) in \cite{Chao:2011cy} reads
\be
\label{psi^1_p_RTA}
\psi^1_p = -\frac{\lambda m}{2}
  \left( \bar{c}^{ij}\sigma_{ij} -\left( c^2 -\alpha_1\right)
    c^i q_i \right) .
\ee
The pre-factor in eq.~(\ref{psi1_all}) can be viewed as an 
explicit result for the relaxation time $\lambda$, which incorporates
the expectation $\lambda = h(z)/T$ which follows from scale
invariance. Scale invariance does not fix the function $h(z)$. 
The factor $\alpha_0$ is equal to one in the RTA. This is 
a shortcoming of that model and implies that the RTA gives
the wrong result for the Prandtl number, the dimensionless ratio
of the shear viscosity to the thermal conductivity. 

 We write $\psi^1_p$ as 
\be 
  \psi^1_p= \frac{\bar{a}_0m}{zT}\, \chi^1_p\, , \hspace{1cm}
  \chi^1_p = \left[ \bar{c}^{ij}\sigma_{ij} 
      -\alpha_0 \left( c^2 -\alpha_1\right) c^i q_i \right]\, , 
\ee
so that $f_p^1=(\bar{a}_0m) \bar{f}_p^0\chi^1_p/T^2$ with 
$\bar{f}_p^0=z^{-1}f_p^0=\exp(-\beta c^2)$ and $\beta=m/(2T)$.
We can then split $X_p^1$ into three terms, $[X^1_p]=[X^2_p]^{(1)}
+ [X^2_p]^{(2)}+[X^2_p]^{(3)}$, corresponding to the different
factors in $f^1_p$ that the comoving derivative is acting on. 
We define 
\bea 
 \left[X^1_p\right]^{(1)} &=& \frac{\bar{a}_0 m}{zT}
    \,\chi^1_p\,{\cal D} \,\ln(\bar{f}^0_p)\, , \\
  \left[X^1_p\right]^{(2)} &=& -\frac{2\bar{a}_0 m}{zT}
    \,\chi^1_p\,{\cal D} \,\ln(T)  \, , \\
 \left[X^1_p\right]^{(3)} &=& \frac{\bar{a}_0 m}{zT}
   \, {\cal D}\, \chi^1_p\, . 
\eea   
The first term requires the calculation of ${\cal D}
\ln (\bar{f}_p^0)=-{\cal D}(\beta c^2)$. In order to
express the results in terms of hydrodynamic variables
we use the results collected in App.~\ref{sec:formulas}.
We find
{\allowdisplaybreaks
\bea
&[X_p^1]^{(1)}_{\it orth}  &= \frac{\bar{a}_0m}{zT}\bigg\{
    \beta\bigg[ c^{i}c^{j}c^{k}c^{l}\sigma_{ij}\sigma_{kl}
             -\frac{2}{15} c^{4} \sigma^{2}\bigg]
    + \bigg[ c^{i}c^{j}
        -\frac{1}{3}c^{2}\delta^{ij}\bigg]
         \bigg( \beta\alpha_0\alpha_1^2 q_iq_j 
         - \alpha_0\alpha_1q_i\nabla_j\alpha \bigg)
    \nonumber\\
    && \mbox{}
    +\bigg[ c^{4}c^{i}c^{j}-\frac{c^{6}\delta^{ij}}{3}\bigg]   
        \beta\alpha_0\, q_iq_j
    + \bigg[c^{2}c^{i}c^{j}
        - \frac{1}{3}c^{4}\delta^{ij}\bigg]
         \bigg( -2\beta\alpha_0\alpha_1 q_iq_j 
           +\alpha_0 q_i\nabla_j\alpha\bigg)
    \nonumber\\[0.3cm] 
    && \mbox{}
    + \bigg[\beta c^{i}c^{j}c^{k} - c^{i}\delta^{jk} \bigg]
        \bigg( \alpha_1(\alpha_0+1) \sigma_{ij}q_k
            - \beta^{-1} \sigma_{ij}\nabla_k\alpha\bigg)
    \nonumber\\[0.3cm]
    && \mbox{}
    - \left[\beta c^{2}c^{i}c^{j}c^{k}
        - \frac{7\alpha_{1}}{5}c^{i}\delta^{jk} \right]
         (\alpha_0+1) \sigma_{ij}q_{k}
    \bigg\}\, .
\label{X_p^1_orth}
\eea}
Here, the subscript ${\it orth}$ refers to the fact that we
have added terms to make the result manifestly orthogonal to 
the zero modes. This requires a number of counterterms, which 
are given by the trace terms in the square brackets, see
App.~\ref{sec:formulas}. We write $[X^1_p]=[X^1_p]_{\it orth}
+[X^1_p]_{\it ct}$. In Sect.~\ref{sec:ct} we will show that 
the sum of the counterterms and the corrections contained 
in $\Delta X_p^0$ do not contribute to the conserved currents. 
In App.~\ref{sec:units} we list the physical dimensions of the 
various terms in $[X^1_p]$. Using these results we can check 
that all terms are consistent with $[X_p^1]\sim T^2$.

 The second term is given by 
\bea
& [X_p^1]^{(2)} &= -\frac{2\bar{a}_0m}{zT}
 \left[ \bar{c}^{ij}\sigma_{ij} -\alpha_0 \left( c^2 -
 \alpha_1\right)
    c^i q_i \right]
    \left(  {\cal D}_u\ln(T)- c^k q_k  \right)
\, .
\eea
We can evaluate the comoving derivative of $T$ using the 
Navier-Stokes equation, and add terms to ensure orthogonality
to the zero modes. The result is 
{\allowdisplaybreaks
\bea
& [X_p^1]^{(2)}_{\it orth} &= \frac{2\bar{a}_0m}{zT}\bigg\{
     \left[c^{i}c^{j}\sigma_{ij}
        -\alpha_0\left(c^{2}-\alpha_1\right)c^{k}q_{k} \right]
        \Big(\frac{2}{3}\langle\sigma\rangle\Big)
        \nonumber\\[0.15cm]
    && \mbox{}\;\;
     + \beta^{-1} \Big[\beta c^{i}c^{j}c^{k}\sigma_{ij}q_{k}
       -  c^{i}q_{k}\sigma^{k}_{i}\Big]
     - \alpha_0 \left(c^2-\alpha_1\right)
    \Big[c^{i}c^{j}-\frac{1}{3}\delta^{ij}c^2\Big]q_iq_j
    \bigg\}\, .
\eea}
Again, these terms have the correct dimension. Finally, the
third term is 
\bea
\label{X^1_3}
[X^1_p]^{(3)}_{\it orth}  &=&  \frac{m\bar{a}_0}{zT} 
\bigg\{
 \Big( c^ic^j-\frac{1}{3}\delta^{ij}c^2\Big) 
     \bigg[ \Big({\cal D}_u-\frac{2}{3}
       \langle\sigma\rangle\Big)\sigma_{ij}
         - \sigma^{}_{ik}\sigma_{j}^{\;\;k} 
         - \sigma^{}_{ik}\Omega_{j}^{\;\;k} 
    \nonumber \\
& &  \hspace*{4.5cm} \mbox{}
    + \alpha_0\alpha_1 \nabla_i q_j
    - \alpha_0\beta^{-1}(\nabla_i\alpha) q_j
         \bigg]
   \nonumber \\
& & \hspace{0.6cm}\mbox{} 
   +\Big( \beta c^ic^jc^k-c^{i}\delta^{jk}\Big)
    \, \beta^{-1}\Big[\nabla_k\sigma_{ij} 
            +\alpha_0 \sigma_{ij}q_k \Big]
      \nonumber \\
 & & \hspace{0.6cm}\mbox{}
-\Big( c^2-\alpha_1 \Big) \, c_i \,\alpha_0 \Big[ 
   {\cal D}_uq_i - \langle\sigma\rangle q_i 
     -\frac{1}{2} \sigma_{kj}q_j
     - \frac{1}{2}\Omega_{kj}q_j 
    \Big] \nonumber \\
 & & \hspace{0.6cm}\mbox{} 
 - \Big( c^2c^ic^j-\frac{1}{3}\delta^{ij}c^4\Big) \,
      \alpha_0\,\nabla_i q_j
    \bigg\} \, .
\eea

\section{Second order contributions to the stress tensor and energy
current}
\label{sec:delj2-pi2}

  The results in the previous section contain a sizable number of 
terms. Indeed, we will see that, with a few exceptions, kinetic theory
generates all the terms that are allowed by the symmetries of the
unitary gas. In the following we will arrange these terms into four 
groups. To begin with, we can separate $[X^1_p]$ into terms that 
are even or odd in $c^i$. Even terms contribute to the stress tensor,
and odd terms provide corrections to the energy current. We can 
further distinguish between terms that are non-linear corrections
to the first order terms (terms like $\delta\Pi_{ij}\sim \sigma_{ik}
\sigma^{k}_{\;j}$), and genuinely novel terms that appear at second 
order, such as contributions to the stress tensor that involve 
temperature gradients, $\delta \Pi_{ij}\sim q_iq_j$.

 We first consider terms in the stress tensor that only involve 
gradients of the fluid velocity. These are constructed from 
$\sigma_{ij}$, $\Omega_{ij}$ and comoving derivatives. These
contributions were already computed  in \cite{Schaefer:2014xma}
\bea
\label{X^1_Pi_sh}
[X^1_p]^{\Pi,sh}_{\it orth}  &=&  \frac{m\bar{a}_0}{zT} \left\{
 \frac{m}{2T} \Big( c^ic^jc^kc^l 
     -\frac{2}{15} \delta^{ik}\delta^{jl}c^4\Big)
           \sigma_{ij}\sigma_{kl}\right.
  \nonumber   \\
 & & \hspace{0.6cm}\mbox{} 
    +\Big( c^ic^j-\frac{1}{3}\delta^{ij}c^2\Big) 
     \left.\left[ \Big({\cal D}_u+\frac{2}{3}\langle\sigma\rangle\Big)\sigma_{ij}
         - \sigma^{}_{ik}\sigma_{j}^{\;\;k} 
         - \sigma^{}_{ik}\Omega_{j}^{\;\;k} \right]
   \right\}\, .
\eea
Using eq.~(\ref{pi_ij_2}) and (\ref{C_L_inv}) we 
can determine the contribution to the stress tensor.
We use
\be 
 \delta \Pi_{ij} = \frac{\nu m}{T}
 \left( \frac{2\bar{a}_0}{zT} \right)
   \langle \bar{v}_{ij} | X^1_p\rangle \, , 
\ee
which leads to 
\be
 \delta \Pi_{ij} = \frac{2\sqrt{2}\bar{a}_0^2}{\pi^{3/2}z}\, 
   m^{3/2}T^{1/2} \, 
\left[ \Big({\cal D}_u+\frac{2}{3}\langle\sigma\rangle\Big)\sigma_{ij}
         + \sigma^{}_{\langle ik}\sigma_{j\rangle}^{\;\;k} 
         - \sigma^{}_{\langle ik}\Omega_{j\rangle }^{\;\;k} 
          \right]\, . 
\ee
Next we consider contributions to the stress tensor that involve 
gradients of the temperature and chemical potential. We have 
\bea 
\label{X^1_Pi_th}
[X^1_p]^{\Pi,th}_{\it orth}  &=&  \frac{m\bar{a}_0}{zT}
\Big( c^ic^j-\frac{1}{3}\delta^{ij}c^2\Big) 
\bigg\{ \Big( 
  \big[ \beta\alpha_0\alpha_1^2+2\alpha_0\alpha_1\big] 
   - 2\alpha_0\big[ \beta\alpha_1 + 1\big] c^2
   + \beta\alpha_0\, c^4
\Big) \, q_iq_j \nonumber \\[0.3cm]
 & & \hspace{1.cm}\mbox{}
  + \Big(-\alpha_0\big[ \alpha_1 + \beta^{-1}\big]
    + \alpha_0 c^2
     \Big) \, q_i\nabla_j\alpha 
  + \alpha_0\Big(\alpha_1 - c^2 \Big)
      \,\nabla_i q_j \bigg\}   \, . 
\eea
The corresponding contribution to the stress tensor is 
\be
\delta \Pi_{ij} = \frac{\bar{a}_0^2}{\pi^{3/2}z}\, 
   m^{1/2}T^{3/2}\,\bigg(
       15\sqrt{2} \, q_{\langle i}q_{j\rangle}
      + 2\sqrt{2} \, q_{\langle i}\nabla_{j\rangle}\alpha
      - 6\sqrt{2} \,  \nabla_{\langle i}q_{j\rangle}
   \bigg)  \, .
\ee
The third group of terms consists of contributions to the energy
current that contain comoving derivatives of $q_i$ as well as the 
expansion rate multiplied by $q_i$. We have
\bea 
\label{X^1_j_th}
[X^1_p]^{\jmath,th}_{\it orth}  &=&  -\frac{m\bar{a}_0}{zT}
  \alpha_0 \left( c^2-\alpha_1\right) c^k  
  \Big( {\cal D}_u  +
    \frac{1}{3}\langle\sigma\rangle\Big) q_k\, .     
\eea
The energy current is given by 
\be 
 \delta\jmath_{\epsilon,i} = \frac{\nu}{2mT}
    \langle (C_L)^{-1} (v_ip^2) | X^1_p\rangle 
    = -\frac{2\nu}{zT} \,\alpha_0\bar{a}_0\,
      \langle c_i R_1(x_c) | X^1_p\rangle \, . 
\ee
Here, we have used eq.~(\ref{CL-inv-A1}) as well as 
$\bar{b}_1=-2\alpha_0\bar{a}_0$. This leads to
\be 
   \delta\jmath_{\epsilon,i} = -\frac{45}{2\sqrt{2}\pi^{3/2}}
    \frac{\bar{a}_0^2}{z}\, m^{1/2}T^{3/2}\,
    \Big( {\cal D}_u  +
    \frac{1}{3}\langle\sigma\rangle\Big) q_k\, . 
\ee
Finally, we consider contributions to the energy 
current that involve the shear strain of the fluid. The 
relevant terms are 
\bea
\label{X^1_j_sh}
[X^1_p]^{\jmath,sh}_{\it orth}  &=&  \frac{m\bar{a}_0}{zT}
  \bigg\{ \Big( \beta c^ic^jc^k-c^i\delta^{jk}\Big)
        \Big[ (\alpha_0+1)\alpha_1 + 2\beta^{-1} 
             +\beta^{-1}\alpha_0\Big]
         \sigma_{ij}q_k \nonumber \\
  &  &  \hspace{1cm}\mbox{} 
    - \Big( \beta c^2c^ic^jc^k-\frac{7}{5}\alpha_1 c^i\delta^{jk}\Big)
        \left(\alpha_0+1\right)  \sigma_{ij}q_k \nonumber \\
  &  &  \hspace{1cm}\mbox{} 
    + \left(c^2-\alpha_1\right) c_i \, \alpha_0\, \frac{1}{2}\,
        \big[ \sigma_{ij}q_j+\Omega_{ij}q_j \big] \nonumber \\
    &  &  \hspace{1cm}\mbox{}      
         + \Big( \beta c^ic^jc^k-c^i\delta^{jk}\Big)
         \beta^{-1} \big[ - \sigma_{ij}\nabla_k\alpha
           + \nabla_k\sigma_{ij}\big]
     \bigg\}\, . 
\eea
The contribution to the energy current is 
\be 
   \delta\jmath_{\epsilon,i} = 
   - \frac{\bar{a}_0^2}{\pi^{3/2}z}\, m^{1/2}T^{3/2}\,
    \Bigg( \frac{231}{4\sqrt{2}}\, \sigma_{ij}q^j
    + \frac{45}{4\sqrt{2}}\, \Omega_{ij}q^j
    + 3\sqrt{2} \Big[ \sigma_{ij}\nabla^j\alpha
       -\nabla^j\sigma_{ij} \Big]
     \Bigg)\, . 
\ee
We can compare these results to the most general form of the 
stress tensor and energy current given in eq.~(\ref{del_pi_fin}) 
and (\ref{del_j_fin}). We observe that there are three second 
order structures that are allowed by the symmetries, but do not 
appear in kinetic theory. These structures are $\Omega_{i}^{\,k}
\Omega_{kj}$, $\nabla_i\alpha\nabla_j\alpha$ and $\nabla_i
\nabla_j\alpha$.

\section{Second order solution: Collision terms}
\label{sec:coll}

The second order collision operator is 
\be
  C^{2}_L\left[\psi^1_1\right] =
  -\int d\Gamma_{234}\, w(1,2;3,4)\, \frac{f^0_2}{T}\,
      \left( \psi^1_1\psi^1_2-\psi^1_3\psi^1_4 \right) \, ,
\ee
where $d\Gamma_{234}=d\Gamma_2\,d\Gamma_3\, d\Gamma_4$, and 
we have used a more compact notation $\psi^1_i\equiv \psi^1_{p_i}$.
Writing $\psi^1_p=(\psi^1_p)_{\it sh}+(\psi^1_p)_{\it th}$ we 
observe that there are three contribution, which we will denote 
as shear-shear (ss), shear-thermal (st), and thermal-thermal (tt).
From the symmetries of the collision integral we can see that the 
stress tensor receives contributions from the shear-shear and
thermal-thermal terms, and that the shear-thermal part contributes
to the energy current. We will write
\be
 C^2_L = (C^2_L)^{\it ss} + (C^2_L)^{\it st}
    + (C^2_L)^{\it tt} \, .  
\ee

\subsection{Second order collision term: Shear-shear}
\label{sec:coll_sh}

 We begin with the shear-shear term. This term was calculated in 
\cite{Schaefer:2014xma}, and we can restrict ourselves to briefly
summarizing the result. We have
\be
  (C^{2}_L\left[\psi^1_1\right])^{\it ss}  =
  - \left(\frac{\bar{a}_0m}{zT}\right)^2\, 
      \sigma_{ij}\sigma_{kl}
    \int d\Gamma_{234}\, w(1,2;3,4)\, \frac{f^0_2}{T}\,
     \left(\bar{c}_1^{ij}\bar{c}_2^{kl} 
           - \bar{c}_3^{ij}\bar{c}_4^{kl}\right) \, ,
\ee
where we have used $(\psi^1_p)^{\it sh}=(\bar{a}_0m)(zT)^{-1}
\bar{c}^{ij}\sigma_{ij}$. The stress tensor is determined by 
the matrix element
\be 
 \big \langle \bar{c}_1^{ab} \big|   
           (C^{2}_L\left[\psi^1_1\right])^{\it ss}\big\rangle 
 \equiv  \left(\frac{\bar{a}_0m}{zT}\right)^2\, 
       [(C^2_L)^{\it ss}]^{ab}_{\;\; ijkl}\; 
        \sigma^{ij}\sigma^{kl} \,\, . 
\ee
Here, $[(C^2_L)^{\it ss}]^{ab}_{\;\; ijkl}$ is a rank 6 tensor 
which is symmetric and traceless in the pairs $(ab)$, $(ij)$ and 
$(kl)$, as well as symmetric under the exchange $(ij)\leftrightarrow
(kl)$. These relations fix the tensor structure and we get
\be 
[(C^2_L)^{\it ss}]^{ab}_{\;\; ijkl}\, \sigma^{ij}\sigma^{kl} 
    = \frac{12}{35}\, ({\cal C}^2_L)^{\it ss} \,
    \sigma^{\langle a}_{\;\;\; c\,}\sigma_{}^{b\rangle c} \, ,
\ee
where we have defined the scalar collision integral
$({\cal C}^2_L)^{\it ss} \equiv 
[(C^2_L)^{\it ss}]^{ab\;\;c}_{\;\;\;\; a\;\; cb}$
given by   
\be
({\cal C}^2_L)^{\it ss} 
 = -\int d\Gamma_{1234}\, w(1,2;3,4)\, \frac{f^0_1f^0_2}{T}\,
     \left(\bar{c}_1\right)^{ab}
     \left[ \left(\bar{c}_1\right)_{a}^{\;\,c}
            \left(\bar{c}_2\right)_{cb} 
          - \left(\bar{c}_3\right)_{a}^{\;\,c}
            \left(\bar{c}_4\right)_{cb}     \right]\, . 
\ee
The scalar collision integral can be computed using the techniques
described in Sect.~\ref{sec:coll_sh_th}. In the unitary limit
$a\to\infty$ we showed that \cite{Schaefer:2014xma}
\be
({\cal C}^2_L)^{\it ss} 
   =   \frac{4T^{11/2}}{9\pi^{5/2}m^{3/2}}\, .
\ee
From this result we obtain the correction to the stress
tensor
\be
\label{c_ij_C^2}
 \delta\Pi_{ij}^2 
  = -\frac{\nu}{2mT}
 \big \langle (C_L^1)^{-1} c_{ij} \big|   
    (C^{2}_L\left[\psi^1_p\right])^{\it ss}\big\rangle 
    = -\frac{64}{105\pi^{5/2}z} \, \bar{a}_0^3
      \, m^{3/2}T^{1/2}\, 
     \sigma_{\langle i}^{\;\;\; k}\sigma^{}_{j\rangle k}\, . 
\ee

\subsection{Second order collision term: Shear-thermal}
\label{sec:coll_sh_th}

 The shear-thermal contribution to the the second order 
collision operator is 
\bea
  (C^{2}_L\left[\psi^1_1\right])^{\it st} &=&
  -\int d\Gamma_{234}\, w(1,2;3,4)\, \frac{f^0_2}{T}\,
      \Big\{ (\psi^1_1)^{\it sh}(\psi^1_2)^{\it th}    
       \nonumber  \\[0.1cm]
   && \mbox{}\hspace{3.cm}     
            +(\psi^1_1)^{\it th}(\psi^1_2)^{\it sh}   
            -(\psi^1_3)^{\it sh}(\psi^1_4)^{\it th} 
            -(\psi^1_3)^{\it th}(\psi^1_4)^{\it sh}
            \Big\} \, , 
\eea
where $(\psi^1_p)^{\it th} = -\bar{b}_1z^{-1} R_1(x_c)c^i 
q_i$. This term contributes to the energy current. The relevant
matrix element is 
\bea
\langle c_i R_1(x_c) | (C_L^2[\psi_p^1])^{\it st}\rangle 
  &=& \frac{m \bar{a}_0 \bar{b}_1 }{z^2 T} 
\int d\Gamma_{1234}\, w(1,2;3,4)\, \frac{f^0_1f^0_2}{T}
  \, (c_1)_i\ R_1(x_1)  \nonumber \\[0.2cm]
  & & \mbox{} \hspace{0.75cm}
  \times \Big[(c_1)_j R_1(x_1) (\bar{c}_2)_{k l} 
             + (c_2)_j R_1(x_2) (\bar{c}_1)_{k l} 
             \nonumber \\[0.1cm]
  & & \mbox{} \hspace{1cm}
             - (c_3)_j R_1(x_3) (\bar{c}_4)_{k l} 
             - (c_4)_j R_1(x_4) (\bar{c}_3)_{k l}\Big] 
             q_j \sigma_{k l}\, , 
\eea
where we have defined $x_i=x_{c_i}$. In this case, the matrix
element of the collision operator defines a rank 4 tensor
\be 
\langle c^i R_1(x_c) | (C_L^2[\psi_p^1])^{\it st}\rangle 
 = -\left( \frac{m\bar{a}_0 \bar{b}_1}{z^2 T}\right) 
    [(C_L^2)^{\it st}]^{ij}_{\;\; k l} \, q_j \sigma^{k l}\, . 
\ee
We can use the symmetries of the problem to express the 
matrix element in terms of a scalar collision integral
\be 
[(C_l^2)^{\it st}]^{ij}_{\;\; kl} \, q_j \sigma^{kl} 
  = \frac{1}{5} ({\cal C}_L^2)^{\it st} q_j \sigma^{ij}\, , 
\ee
where we have defined $({\cal C}_L^2)^{\it st}= 
[(C_l^2)^{\it st}]^{ij}_{\;\; ij}$. The scalar collision
integral can be written as 
\bea
({\cal C}_L^2)^{\it st} &=& 
  -\int d\Gamma_{1234}\, w(1,2;3,4)\,  \frac{f^0_1f^0_2}{4T}
   \Big[ 
     (c_1)_i R_1(x_1) + (c_2)_i R_1(x_2)
   - (c_3)_i R_1(x_3) - (c_4)_i R_1(x_4) \Big] 
     \nonumber \\
  & & \Big[
    (c_1)_j R_1(x_1) (\bar{c}_2)^{i j} 
  + (c_2)_j R_1(x_2) (\bar{c}_1)^{i j} 
  - (c_3)_j R_1(x_3) (\bar{c}_4)^{i j} 
  - (c_4)_j R_1(x_4) (\bar{c}_3)^{i j}  \Big]\, , 
\eea
where we have made use of the symmetries of the integrand. 
We compute the integral by introducing center-of-mass and 
relative momenta
\be 
 m\,\vec{c}_{1,2} = \frac{\vec{P}}{2}\pm \vec{q} \, ,\hspace{0.5cm}
 m\,\vec{c}_{3,4} = \frac{\vec{P}}{2}\pm \vec{q}^{\,\prime}\, , 
\ee
and write the measure on phase space as
\begin{align} 
\int d\Gamma_{1234} \, (2\pi)^4 & 
       \delta^3\Big(\sum_i \vec{p}_i\Big)
    \delta \Big(\sum_i E_i\Big)\nonumber \\
  & = \frac{2}{(2\pi)^6} \int P^2dP \int q^2 dq \,\frac{qm}{2} \, 
      \int d\cos\theta_q\int d\cos\theta_{q'}\int d\phi_{q'}\, . 
\end{align}
Here we have chosen a coordinate system in which $\vec{P}=P\hat{z}$,
so that $\hat{P}\cdot\hat{q}=\cos\theta_{q}$. We also use $\hat{P}
\cdot\hat{q}'=\cos\theta_{q'}$ as well as $\hat{q}\cdot\hat{q}'
=\cos\theta_{q}\cos\theta_{q'}-\sin\theta_{q}\sin\theta_{q'}
\cos\phi_{q'}$. We observe that neither the product of the 
distribution functions, $f_1^0f_2^0$, nor the scattering 
amplitude, $|{\cal A}|^2$, depend on the angles $\theta_q,
\theta_{q'}$ and $\phi_{q'}$. We obtain
\bea 
({\cal C}_L^2)^{\it st} &=&  -\frac{z^2}{(2 \pi)^6T} 
  \int P^2 dP\, \int q^2 dq\, \int d\cos\theta_q\,\int  d\cos\theta_{q'}
        \int d\phi_{q'}\,  \nonumber \\
  & & \mbox{} \hspace{3.25cm} \times
       \left(\frac{16 \pi ^2}{m q}\right) 
  \exp\left(-\frac{P^2}{4 m T}\right)
  \exp\left(-\frac{q^2}{m T}\right) \, 
  \Omega_{qq'}
\eea
where the angular factor is given by 
\bea 
\Omega_{qq'} &=& 
  \frac{P^2 q^4}{48 m^6 T^2} 
     \big[ 2\cos(\theta_q) \cos(\theta_{q'}) 
     \left(20 m T + 3 P^2 - 12 q^2 \right) \nonumber \\
  & & \mbox{} \hspace{2.5cm}   
     \big( \cos(\theta_q) \cos(\theta_{q'}) 
           - \sin(\theta_q) \sin(\theta_{q'}) \cos(\phi_{q'}) \big) 
            \nonumber \\
  & & \mbox{}
     + \cos^2(\theta_q)\big(12 q^2 - 20 m T + P^2 \cos(2\theta_q) 
     - P^2\cos(2\theta_{q'}) - 3 P^2\big) 
        \nonumber \\
& & \mbox{}
    + \cos^2(\theta_{q'}) \big(12 q^2 - 20 m T - P^2 \cos(2\theta_q) 
    + P^2\cos(2\theta_{q'}) - 3 P^2\big)
\big]\, . 
\eea
The integrals can be carried out and we find 
\be 
\label{C_L_2-st-res}
({\cal C}_L^2)^{\it st} = 
  \frac{4 z^2 T^{7/2}} {3\pi^{5/2} m^{1/2}}\, . 
\ee
The contribution to the energy current is 
\be 
\delta \jmath_{\epsilon,i}^2 = 
  -\frac{\nu}{2 m T} \langle C_L^{-1} (v_i p^2) | 
       C_L^2[\psi_p^1]\rangle
     = -\frac{\nu \bar{b}_1}{zT} 
           \langle c_i R_1 (x_c) | C_L^2[\psi_p^1]\rangle \, , 
\ee
from which we finally obtain 
\be 
\label{del-j-coll-st}
\delta \jmath_{\epsilon,i}^2 
   = \frac{24}{5\pi^{5/2}z} 
   \,  \bar{a}_0^3\, m^{1/2} T^{3/2}\,  q^j \sigma_{i j} \,  . 
\ee

\subsection{Second order collision term: Thermal-thermal}
\label{sec:coll_th_th}

The thermal-thermal part of the second order collision term
contributes to the stress tensor. The relevant matrix element 
is 
\bea
\langle \bar{c}_{ij} | (C_L^2[\psi_p^1])^{\it tt}\rangle 
  &=& -  \left(\frac{\bar{b}_1}{z}\right)^2 
\int d\Gamma_{1234}\, w(1,2;3,4)\, \frac{f^0_1f^0_2}{T}
  \, (c_1)_{ij}  \nonumber \\[0.2cm]
  & & \mbox{} \hspace{0.75cm}
  \times \Big[(c_1)_k (c_2)_l R_1(x_1)R_1(x_2)
            - (c_3)_k (c_4)_l R_1(x_3)R_1(x_4) \Big] 
           \;  q^k q^l\, . 
\eea
Again, the matrix element of the collision operator defines
a rank 4 tensor
\be 
\langle \bar{c}^{ij} | (C_L^2[\psi_p^1])^{\it tt}\rangle 
 = \left( \frac{\bar{b}_1}{z}\right)^2 
    [(C_L^2)^{\it tt}]^{ij}_{\;\; k l} \, q_k q^l \, , 
\ee
and symmetries can be used to write the matrix element in 
terms of a scalar collision integral
\be 
[(C_l^2)^{\it tt}]^{ij}_{\;\; kl} \, q^k q^l
  = \frac{1}{5} ({\cal C}_L^2)^{\it tt} 
         q^{\langle i}q^{j\rangle} \, , 
\ee
where we have defined $({\cal C}_L^2)^{\it tt} \equiv
[(C_l^2)^{\it tt}]^{ij}_{\;\; ij}$. The scalar collision
integral is
\bea
({\cal C}_L^2)^{\it tt} &=& 
  -\int d\Gamma_{1234}\, w(1,2;3,4)\,  \frac{f^0_1f^0_2}{8T}
   \Big[ 
     (c_1)^{ij} + (c_2)^{ij} 
   - (c_3)^{ij} - (c_4)^{ij} \Big] 
     \nonumber \\[0.1cm]
  & & \mbox{}\hspace{1.5cm}
  \times\Big[ \big\{ (c_1)_i(c_2)_j 
                   + (c_2)_i(c_1)_j \big\}
                       R_1(x_1)R_1(x_2)
     \nonumber \\[0.2cm]
 & & \mbox{}\hspace{1.75cm}   
      - \big\{ (c_3)_i (c_4)_j 
             + (c_4)_i (c_4)_j \big\}
                      R_1(x_3)R_1(x_4)
   \Big]\, , 
\eea
where we have made use of the symmetries of the integrand. 
The integrals can be carried out analytically and we obtain
\be 
\label{C_L_2-tt-res}
({\cal C}_L^2)^{\it tt} = 
  \frac{z^2 T^{7/2}}{2\pi^{5/2}m^{1/2}}\, ,
\ee
which determines the contribution to the stress tensor
\be 
\label{del-pi-coll-tt}
\delta \Pi^2_{ij} 
   = -\frac{18}{5 \pi ^{5/2} z} \, \bar{a}_0^3 \, m^{1/2} T^{3/2}  
   \, q_{\langle i} q_{j\rangle} \,  . 
\ee

\section{Counterterms}
\label{sec:ct}

 The matrix elements that determine the stress tensor and
energy current, given in eq.~(\ref{pi_ij_2}) and (\ref{delta-j2}),
contain the expression
\bea 
\label{del-X1p-def}
 \Delta X_p^0 + X^1_p - C_L^2[\psi^1_p]
   &=&   \left( [X^1_p]_{\it orth}- C_L^2[\psi^1_p] \right) +
   \left(  \Delta X_p^0 + [X^1_p]_{\it ct}\right) 
    \nonumber \\
    &\equiv&   \; [X^1_p]_{\it hydro}  + \delta X^1_p  \, ,
\eea
where we have used $[X^1_p]=[X_p^1]_{\it orth}+[X^1_p]_{\it ct}$, 
see the discussion below eq.~(\ref{X_p^1_orth}). Note that 
$[X^1_p]_{\it hydro}$ is orthogonal to the zero modes of the
collision operator, and that it contributes to the stress tensor 
and energy current at second order in the gradient expansion. We 
have computed these contribution in Sec.~\ref{sec:delj2-pi2} and 
\ref{sec:coll}. The terms in $\delta X^1_p$ are not manifestly 
orthogonal to the zero modes, and it is not immediately clear 
whether they contribute to the conserved currents. In this 
section we show that $\delta X^1_p$ is indeed orthogonal to 
the zero modes, and that it does not contribute to the conserved 
currents. We can view $\delta X^1_p$ as a set of terms in kinetic 
theory that contribute to non-hydrodynamic moments of the 
distribution function. 

 We first compute $\Delta X_p^0= X_p^0-C_L^1[\psi^1_p]$. As
explained above, $\Delta X^0_p$ vanishes at first order in the 
gradient expansion. Non-zero contributions to $\Delta X^0_p$
arise when we use the Navier-Stokes equation rather than the 
Euler equation in order to simplify $X_p^0$. We can write the 
Navier-Stokes equation as 
\bea 
\label{ns_d_1}
 {\cal D}_u\alpha\;  &=&  (\delta^2 \alpha)\, , \\
\label{ns_d_2}
 {\cal D}_u u_i   \; &=& - \frac{T}{m} \left( \nabla_i\alpha  
   +\frac{5}{2} \nabla_i\ln(T) \right)
   + (\delta^2 u_i)
   \, , \\
\label{ns_d_3}
 {\cal D}_u\ln(T)   &=&  - \frac{2}{3}\langle \sigma\rangle 
   + (\delta^2 T)   \, ,
\eea
where $(\delta^2\alpha),(\delta^2 u_i)$ and $(\delta^2 T)$ are
terms of second order in gradients which are computed in
App.~\ref{sec:NS}, see eqs.~(\ref{ns_1}-\ref{ns_3}). In 
particular, we find that $(\delta^2 T)$ and $(\delta^2 \alpha)$
contain $\sigma^2\equiv\sigma_{ij}\sigma^{ij}$, $q^2$ and 
$\nabla\cdot q$, and $(\delta^2 u_i)$ is proportional to a 
linear combination of $\nabla_j\sigma_{ij}$ and $q_j\sigma_{ij}$. 
Using the result for $X_p^0$ given in eq.~(\ref{X^0}) together 
with the Navier-Stokes equation we find
\be 
\label{del-Xp0}
\Delta X_p^0 = \frac{m}{2}\left\{ c^2 (\delta^2 T) 
  + 2c^i (\delta^2 u_i) + \frac{2T}{m} (\delta^2\alpha) 
  \right\}\, . 
\ee  
The counterterms in $[X^1_p]_{\it ct}$ are the terms 
that were introduced in Sect.~\ref{sec:X1} in order to make 
the second order streaming terms manifestly orthogonal to the
zero modes. We can group these terms into three categories, 
pure shear terms, pure heat flow, and terms that involve 
combinations of shear stress and heat flow. The pure shear 
terms are
\bea
\label{X1p-ct-sh}
[X^1_p]_{\it ct} &=& \frac{\bar{a}_0m}{zT}
 \left\{  
   \left[ \frac{2}{15}\, \beta c^4 
            - \frac{1}{3}\, c^2\right]\, \sigma^2
        + c^i \beta^{-1} \nabla^j \sigma_{ij}
   \right\}\, .
\eea
Pure thermal conduction terms are given by 
\bea
[X^1_p]_{\it ct} &=& \frac{\bar{a}_0m}{zT}
 \left\{  
   \left[ \frac{\beta\alpha_0}{3} c^6
        - \frac{2}{3}\big(\beta\alpha_0\alpha_1 +\alpha_0\big)c^4
        + \left(\frac{\beta\alpha_0\alpha_1^2}{3}
             + \frac{2\alpha_0\alpha_1}{3}   
             - \frac{5\alpha_0}{4\beta}\right)c^2 
    \right.\right.  \nonumber \\
 & & \hspace{1.55cm}\mbox{}\left.\left. 
             - \frac{5\alpha_0\alpha_1}{4\beta}
                  \right]\, q^2 
     +\left[ -\frac{\alpha_0}{3} c^4 
              + \frac{\alpha_0\alpha_1}{3} c^2\right]
        \nabla\cdot  q
   \right\}\, .
\label{X1p-ct-th}
\eea
Finally, the mixed terms are
\bea
[X^1_p]_{\it ct} &=& \frac{\bar{a}_0m}{zT}
 \left\{  
    \left[  \frac{\alpha_0}{3} c^4 
        - \left(\frac{\alpha_0\alpha_1}{3}
                   +\frac{5\alpha_0}{6\beta} \right) c^2
            +\frac{\alpha_0\alpha_1}{2\beta}
         \right] q\cdot\nabla\alpha \right.\nonumber \\
  & & \hspace{1cm}\mbox{}\left. 
 -  \left[ \frac{2(\alpha_0+1)\alpha_1}{5} - \frac{1}{\beta}
    \right] c^iq^j\sigma_{ij}
 \right\}\, . 
 \label{X1p-ct-mix}
\eea 
There is one more allowed structure, proportional to $\sigma_{ij}
\nabla^j\alpha$, which has a vanishing coefficient. We can 
now compute $\delta X^1_p$ as the sum of 
eqs.~(\ref{del-Xp0}-\ref{X1p-ct-mix}), and determine the 
contribution to the conserved densities and currents. We find
\be 
  \langle \delta X^1_p|\chi^{(k)}_{\it zm}\rangle = 0 
\ee
for $\chi^{(k)}_{\it zm}=\{1,c_j,c^2\}$, as well as 
\be 
  \langle \delta X^1_p|c^ic^j\rangle = 
  \langle \delta X^1_p|c^ic^2\rangle = 
  0 \, .
\ee

\section{Second order transport coefficients}
\label{sec:res}

We can read off the transport coefficients by matching the 
second order corrections to the currents to the general result in 
conformal fluid dynamics, eqs.~(\ref{del_pi_fin},\ref{del_j_fin}).
For this purpose we combine the results in Sect.~\ref{sec:delj2-pi2} 
with the collision integrals computed in Sect.~\ref{sec:coll}.
We previously determined shear contributions to the stress
tensor \cite{Schaefer:2014xma}. These terms are
\be 
\label{final}
\tau_\pi=\frac{\eta}{P}\, , \hspace{0.3cm}
\lambda_1 =  \frac{15}{14}\frac{\eta^2}{P}\, , \hspace{0.3cm}
\lambda_2 = -\frac{\eta^2}{P}\, , \hspace{0.3cm}
\lambda_3 = 0 \, ,
\ee
and $\Theta_\pi=2/3$. Here we have used the kinetic theory result for 
$\eta$ and $P$, see eq.~(\ref{kin-e}) and (\ref{kin-eta}). Transport 
coefficients that determine the stress tensor in the presence of thermal 
gradients are
\be
\label{fin_pi_th}
\gamma_1=\frac{169}{120} \frac{m\kappa^2}{c_P} \, ,  \hspace{0.3cm}
\gamma_2= \frac{8}{45} \frac{m\kappa^2}{c_P}\, ,   \hspace{0.3cm}
\gamma_3=0  \,  , \hspace{0.3cm}
\gamma_4= -\frac{8}{15} \frac{m\kappa^2}{c_P} \, ,   \hspace{0.3cm}
\gamma_5=0   \, .
\ee
The result for $\gamma_1$ receives contributions from both the 
second order collision term and the streaming terms, whereas 
$\gamma_2$ and $\gamma_4$ are determined by the streaming terms
only. The coefficients $\gamma_3$ and $\gamma_5$ vanish because
the distribution function at first order does not contain terms
proportional to $\nabla_i\alpha$. As a result, kinetic theory
does not generate second order terms of the form $\nabla_i\nabla_j
\alpha$ or $(\nabla_i\alpha)(\nabla_j\alpha)$. The transport 
coefficients that determine the energy current are given by  
\be 
\label{fin_j_th}
\tau_\kappa=\frac{m\kappa}{c_PT}\, , \hspace{0.3cm}
\nu_1 = -10\,\frac{\eta\kappa}{c_P}  \, , \hspace{0.3cm}
\nu_2 =-\frac{15}{8}\frac{\eta\kappa}{c_P}  \, , \hspace{0.3cm}
\ee
with $\Theta_\kappa=1/3$, and
\be
\label{fin_j_oth}
\nu_3= -\frac{\eta\kappa}{c_P}  \, , \hspace{0.3cm}
\nu_4= 0\, ,  \hspace{0.3cm}
\nu_5= \frac{\eta\kappa}{c_P}  \, , \hspace{0.3cm}
\nu_6 =0\, . 
\ee
Again, $\nu_1$ is the most complicated term, because it 
receives contributions from both the streaming terms and 
the collision terms. The coefficients $\nu_4$ and $\nu_6$
vanish, because the first order result contains neither
$\nabla_i\alpha$ nor $\Omega_{ij}$.

\section{Conclusions and outlook}
\label{sec:final}

In this work we have determined the complete set of second order 
transport coefficients of the unitary Fermi gas in kinetic theory. 
It will be difficult to measure all of them individually, but the 
relaxation times as well as certain linear combinations of the 
transport terms are more readily accessible. Li et al.~\cite{Li:2024} 
reported a measurement of thermal and viscous relaxation rates, in 
rough agreement with the Boltzmann theory \cite{Frank:2020}. The 
experiment is based on the evolution of a density perturbation
after the perturbing potential is turned off. Li et al.~noted 
that a density perturbation at constant temperature corresponds 
to a linear combination of a sound mode and a diffusive heat mode. 
As a result, the decay of the initial perturbation contains 
an underdamped mode sensitive to the sound diffusivity, and an
overdamped mode controlled by thermal conductivity. The early-time
evolution of that mode is sensitive to thermal relaxation.

 In the present work we confirmed the result $\tau_\kappa=m\kappa/
(c_PT)$ in a calculation that includes the full second order 
collision term. Note that this result is straightforward to derive 
in the relaxation time (RTA) approximation, but the RTA calculation 
gives a result for $\kappa$ that is off by a factor 2/3. 
Another quantity that is experimentally accessible is the 
second order contribution to the sound attenuation rate, which
leads to a nonlinear dependence of the damping rate on the 
square of the wave number. We have given a rough estimate of 
the magnitude of this effect in Sect.~\ref{sec_scales}, 
where we have neglected thermal conduction and non-linear 
effects in the velocity field. We will study this issue in 
more detail in a forthcoming publication. 

 Acknowledgments: This work was supported in parts by the US Department 
of Energy grant DE-FG02-03ER41260 and DE-SC0024622. We thank John 
Thomas for many useful discussions. 

\appendix

\section{Definitions and units}
\label{sec:units}

  We first summarize various definitions that are made in the 
text. The fluid velocity is denoted by $u^k$, the velocity of 
a particle is $v^k=p^k/m$, and the relative velocity is $c^k=v^k-u^k$.
The thermal gradient is $q^k=-\nabla^k\ln(T)$. The comoving 
derivatives are 
\be 
{\cal D}_u=\partial_t +  u^k\nabla_k 
 \;\;\; ({\it fluid})\, ,  \;\;\;\;\;\;
 {\cal D}=\partial_t +  v^k\nabla_k \,
 \;\;\; ({\it particle})\, . 
\ee
The inverse temperature and fugacity are
\be 
\beta=\frac{m}{2T}\, , \hspace{1cm}
 z=e^\alpha = e^{\mu/T}\, . 
\ee
Finally, the distribution function contains the two parameters
\be
 \alpha_1=\frac{5T}{m} \, , \hspace{1cm}
 \alpha_0=\frac{3}{2}\, . 
\ee
In order to check the consistence of our results it is useful
to tabulate the physical units of various quantities.
We set Boltzmann's constant and Planck's constant to one, 
and use mass $m$ and temperature $T$ as the basic units. 
We use $[O]$ to denote the physical units of an observable 
$O$. We have
\be  
 [\nabla]= \sqrt{mT}\, , \hspace{1cm}
 [{\cal D}]=[\partial_t]= T\, ,
\ee
\be
 [c^2]= \frac{T}{m} \, , \hspace{1cm}
 [\sigma]=[\nabla\cdot u] = T \, , \hspace{1cm}
 [q]= \sqrt{mT}\, , \hspace{1cm} 
 [\alpha]=1\, , 
\ee
\be 
[\beta] = \frac{m}{T}\, , \hspace{1cm}
[\alpha_0]= 1\, , \hspace{1cm}
[\alpha_1]= \frac{T}{m}\, .
\ee
The distribution functions have units
\be
[f] =1\, , \hspace{1cm}
[\psi] = T\, , \hspace{1cm}
[X] = T^2\, .
\ee
The collision integrals and the linearized collision
operator have units
\be
 [C_L^1[\psi_p^1]] = [ C_L^2[\psi_p^1]] = T^2\, , \hspace{1cm}
 [(C_L^1)] = T\, . 
\ee
The currents have units
\be 
[\jmath^\varepsilon_i ] =mT^3\, , \hspace{1cm}
[\Pi_{ij}] = m^{3/2}T^{5/2}\, , 
\ee
and the first order transport coefficients have units
\be 
[\eta]= (mT)^{3/2}\, , \hspace{1cm}
[\kappa] = m^{1/2}T^{3/2}\, . 
\ee
Finally, the physical units of the second order transport 
coefficients are
\be 
[\lambda_i]= m^{3/2}T^{1/2}\, , \hspace{0.75cm}
[\gamma_i] = m^{1/2}T^{3/2}\, , \hspace{0.75cm}
[\nu_i]    = m^{1/2}T^{3/2}\, . 
\ee

\section{Navier-Stokes equation}
\label{sec:NS}

  The Navier-Stokes equation corresponds to the conservation
laws for mass, momentum and energy, where the corresponding 
currents have been truncated at the one-derivative level, see 
eqs.~(\ref{hydro-1}-\ref{q_i}). In the following, we will rewrite 
these equations in terms of the variables $\alpha, \ln(T)$ and 
$u_i$. For this purpose we will use the thermodynamic identities
\be
\label{Gibbs-Dukem}
dP=nd\mu + sdT = nTd\alpha + (n\mu +sT)d\ln(T)
\ee
and ${\cal E}+P=n\mu + sT$. We will also use kinetic theory.
At leading order in the fugacity we get
\be
{\cal E}= \frac{3}{2} nT\, , \hspace{0.75cm}
  P = nT\, , \hspace{0.75cm}
  c_p=\frac{5}{2}\, n \, , \hspace{0.75cm}
  n = \nu z \left(\frac{mT}{2\pi}\right)^{3/2} \, ,
  \label{kin-e}
\ee  
as well as 
\be
  \eta = - 4\bar{a}_0\, 
       \left(\frac{mT}{2\pi}\right)^{3/2} , 
  \hspace{0.5cm}
  \kappa = \;\,\frac{5\bar{b}_1}{m}\, 
       \left(\frac{mT}{2\pi}\right)^{3/2} ,
\label{kin-eta}
\ee
with $\bar{b}_1=-2\alpha_0\bar{a}_0$. Here, we have made use
of the kinetic theory results for $\eta$ and $\kappa$ derived
earlier. Eq.~(\ref{kin-eta}) implies that $\eta/n=-2\bar{a}_0/z$ 
and $\kappa/n=-5\alpha_0\bar{a}_0/(mz)$. We then get 
\bea 
\label{ns_1}
 {\cal D}_u\alpha\;  &=& 
    \frac{15\alpha_0\bar{a}_0}{2mz}
    \left(-\frac{2}{3}\nabla_k q^k + \frac{5}{2} q^2\right)
      + \frac{\bar{a}_0}{zT} \,\sigma^2\\
\label{ns_2}
 {\cal D}_u u_i   \; &=& - \frac{T}{m} \left( \nabla_i\alpha  
   +\frac{5}{2} \nabla_i\ln(T) \right)
   -\frac{2\bar{a}_0}{mz} \left(\nabla^j\sigma_{ij} 
    -\frac{3}{2}\sigma_{ij}q^j \right)
   \, , \\
\label{ns_3}
 {\cal D}_u\ln(T)   &=&  - \frac{2}{3}\langle \sigma\rangle 
   - \frac{5\alpha_0\bar{a}_0}{mz}
    \left(-\frac{2}{3}\nabla_k q^k + \frac{5}{2} q^2\right)
      - \frac{2\bar{a}_0}{3zT} \,\sigma^2    \, .
\eea

\section{Kinetic theory formulas}
\label{sec:formulas}

 Here we collect a number of simple relationships that are used
in order to simplify the streaming terms. First, we observe that 
there is a simple relationship between the comoving derivatives 
in the kinetic theory and in fluid dynamics, 
\be 
{\cal D}={\cal D}_u + c^k\nabla_k \, . 
\ee
We also observe that 
\be
 {\cal D}_u c_i = -{\cal D}_u u_i, \hspace{0.7cm}
 \nabla_k c_i = -\nabla_k u_i\, , \hspace{0.7cm}
 {\cal D} c_i = -({\cal D}_u+c^k\nabla_k) u_i\, . 
\ee 
Derivatives of the fluid velocity can be written as 
\be 
\nabla_k u_i = \frac{1}{2}\sigma_{ki} + \frac{1}{2}\Omega_{ki}
   + \frac{1}{3}\delta_{ki} \langle \sigma\rangle\, , 
\ee
where $\sigma_{ki}$ is the shear stress tensor, $\Omega_{ki}$ is
the vorticity tensor, and $\langle\sigma\rangle$ is the bulk stress,
see eq.~({\ref{sig_ij},\ref{del_pi_fin}). Note that comoving 
derivatives of thermodynamics variables such as ${\cal D}_u u_i, 
{\cal D}_u \ln(T)$ and ${\cal D}_u \alpha$ 
can be written in terms of spatial derivative terms 
$\sigma_{ij},q_i$ and $\nabla_i\alpha$ using the Navier-Stokes
equation (\ref{ns_1}-\ref{ns_3}). At first order in derivatives
\be
\label{ns_lin}
 {\cal D}_u\alpha = 0, \hspace{0.5cm}
 {\cal D}_u u_i   = - \frac{T}{m} \left( \nabla_i\alpha  
   - \frac{5}{2} q_i \right), \hspace{0.5cm}
 {\cal D}_u\ln(T)=  - \frac{2}{3}\langle \sigma\rangle .
\ee
The coefficient $\alpha_0$ is a constant and ${\cal D}\alpha_0=
{\cal D}_u\alpha_0=0$. The comoving derivative of $\alpha_1$ is 
\be 
{\cal D} \alpha_1 = -\alpha_1\left( \frac{2}{3} \langle\sigma\rangle
  + c^kq_k \right).
\ee
We employ the following integrals
\bea
 \int d^3c\; c_ic_j f(c^2)
 & = & \frac{1}{3}\, \delta_{ij}
   \int d^3c\; c^2\, f(c^2) \, ,  \\  
   \int d^3c\; c_ic_jc_kc_l f(c^2)
 & = & \frac{1}{15}\, \left(\delta_{ij}\delta_{kl}
   +\delta_{ik}\delta_{jl}+\delta_{il}\delta_{jk}\right)
   \int d^3c\; c^4\, f(c^2) \, , \\
\int d^3c\; c_ac_bc_ic_jc_kc_l f(c^2)
 & = & \frac{1}{105}\, \left[\delta_{ab} \left(\delta_{ij}\delta_{kl}
   +\delta_{ik}\delta_{jl}+\delta_{il}\delta_{jk}\right)
   + {\it perm} \right]
   \int d^3c\; c^6\, f(c^2) \, ,
 \eea
where $f(c^2)$ is an arbitrary function. We also have
\bea
\int d^3c\, \bar{c}_{ij}\bar{c}_{ab} f(c^2)
  =  \frac{1}{15}\left( \delta_{ai}\delta_{bj}
    + \delta_{aj}\delta_{bi} 
    - \frac{2}{3}\delta_{ij}\delta_{ab}\right)
    \int d^3c\, c^4 f(c^2) \, .
\eea
These results can be used in order to make expressions 
involving tensors built from $c_i$ orthogonal to the zero 
modes of the collision operator $\{1,c_i,c_ic^2\}$. We 
obtain
\bea
c_ic_j c^{2n} &\to & 
    \Big( c_ic_j-\frac{1}{3}\delta_{ij}c^2\Big) c^{2n}\, ,  \\ 
c_ic_jc_kc_l c^{2n} &\to & 
    \Big( c_ic_jc_kc_l
    -\frac{1}{15}\delta_{(ij}\delta_{kl)}c^4\Big) c^{2n}\, , \\ 
c_i c^2       &\to & 
    c_i\Big( c^2-\alpha_1\Big)\, ,  \\
\beta c_{\langle i}c_{j\rangle}c_k     &\to & 
    \Big( \beta c_{\langle i}c_{j\rangle}c_k - 
       c_{\langle i}\delta_{j\rangle k} \Big)\, ,  \\
\beta c_{\langle i}c_{j\rangle}c_kc^2     &\to & 
  \Big( \beta c_{\langle i}c_{j\rangle}c_k c^2 
   - \frac{7}{5}\alpha_1 c_{\langle i}\delta_{j\rangle k} \Big) \, ,
\eea
where $\delta_{(ab}\delta_{cd)}$ is completely symmetric in 
all indices, and $c_{\langle i}c_{j\rangle}$ is symmetric 
and traceless. 

\section{Signs}
\label{sec:signs}

 It is useful to qualitatively understand the signs of the
terms in $\delta f$, $\delta\Pi_{ij}$ and $\delta \jmath_i$.
Consider first the first order corrections to the shear stress
and energy current. The first order term in the stress tensor is 
$\delta\Pi_{ij}=-\eta\sigma_{ij}$. Based on the second law of 
thermodynamics, we expect $\eta$ to be positive. We have defined 
$\delta f_p\sim f_p^0\psi^1_p/T$, and we obtained $\psi^1_p \sim
\bar{a}_0\bar{c}^{ij}\sigma_{ij}$. Using the definition of the 
stress tensor in kinetic theory, see eq.~(\ref{pi_ij_kin}), we 
see that $\eta>0$ requires $\bar{a}_0<0$, which is indeed the case,
see eq.~(\ref{a_0_kin}). 

 We can check what this result implies for the distribution function. 
Consider a simple shear flow with $\nabla_y u_x>0$. Then $\delta f
\sim - v_x v_y$, which is an ellipsoidal deformation of the distribution
function in momentum space that increases the number of particles 
with $v_y>0$ and $v_x<0$. This deformation will indeed tend to reduce
a shear flow in which $u_x$ increases with $y$. Note that the force 
on a surface that bounds the fluid at $y=0$ is $F_i=\delta\Pi_{ij}
\hat{n}_j$ with $\hat{n}=-\hat{e}_y$. As expected we get $F_x=\eta
\nabla_y u_x>0$.

 The leading order term in the energy current is $\jmath_i = -\kappa
\nabla_i T$. The second law of thermodynamics states that heat flows 
from hot to cold regions so that $\kappa>0$. The shift in the distribution 
function is $\psi^1_p\sim -\bar{a}_0\alpha_0(c^2-\alpha_1)c^i q_i$, where 
$\bar{a}_a\alpha_0<0$ and $q_i=-\nabla_i \ln(T)$. Consider $\nabla_x T<0$
so that $q_x>0$. Then for particles with $v_x>0$ the distribution 
function is enhanced for large $v^2$, and suppressed for small $v^2$.
This means that on average energy is transported in the positive $x$ 
direction, and the temperature gradient is reduced. 

We conclude that in the result given in eq.~(\ref{psi1_all})
\be
\psi^1_p = \frac{\bar{a}_0m}{zT}
  \left( \bar{c}^{ij}\sigma_{ij} -\alpha_0 \left( c^2 -\alpha_1\right)
    c^i q_i \right) ,
\ee
the signs $\bar{a}_0<0$ as well as $\alpha_{0,1}>0$ are consistent
with physical expectations. 

 The signs of second order contributions are in general
not constrained by the second law. Within kinetic theory 
we expect the relaxation times $\tau_\pi$ and $\tau_\kappa$
that appear in 
\bea 
  \delta\jmath_\epsilon^i &=& \kappa T q^i 
     -\kappa T\tau_\kappa  {\cal D}_u q^i \, ,\\
  \delta \Pi_{ij} &=& -\eta\sigma_{ij} + \eta\tau_\pi
      {\cal D}_u \sigma_{ij}\, , 
\eea
to be positive. This allows the two relations to be 
written (at second order accuracy) as relaxation 
equations
\bea 
  {\cal D}_u \delta\jmath_\epsilon^i &=& 
     -\frac{1}{\tau_\kappa}\left(
   \delta\jmath_\epsilon^i - \kappa T q^i\right)  \, ,\\
   {\cal D}_u \delta\Pi_{ij} &=&
    -\frac{1}{\tau_\pi} \left( 
    \delta \Pi_{ij} +\eta\sigma_{ij} \right)\, , 
\eea
which describe the decay of the currents to the Fourier-Navier-Stokes
form. We do indeed find that $\tau_\pi,\tau_\kappa>0$. This result 
can be traced to the fact that $C_L^1$ has negative eigenvalues. 

Finally, it is interesting to consider the sign of the non-linear terms
in the stress tensor. We find
\be
\Delta\Pi_{ij} = 
\eta\tau_\pi\Theta_\pi \langle \sigma\rangle \sigma_{ij} 
    + \lambda_1 \sigma_{\langle i}^{\;\;\; k}\sigma^{}_{j\rangle k} 
    + \lambda_2 \sigma_{\langle i}^{\;\;\; k}\Omega^{}_{j\rangle k}
    + \lambda_3 \Omega_{\langle i}^{\;\;\; k}\Omega^{}_{j\rangle k}  
    + \gamma_1 q_{\langle i}q_{j\rangle}\, , 
\ee
with $\lambda_1>0$, $\lambda_2<0$, $\lambda_3=0$ and $\gamma_1>0$.
Consider again a pure shear flow $S=\nabla_yu_x>0$. The non-linear
term in the stress tensor is 
\be 
\delta\Pi_{ij} = {\it diag}\Big( (\lambda_1/3+\lambda_2),
                            (\lambda_1/3-\lambda_2),
                            -2\lambda_1/3\Big) S^2 .
\ee
Note that there is no non-linear correction to the tangential force 
$F_x$ on a surface at $y=0$. There is a normal force $F_y = -(\lambda_1
/3-\lambda_2)S^2<0$. In the same fashion we can consider a temperature
gradient $K=\nabla_x T<0$. The stress tensor is 
\be 
\delta\Pi_{ij} = \gamma_1 {\it diag}\Big( 2/3,-1/3,-1/3\Big) K^2/T^2 \, ,
\ee
which gives a positive normal force $F_y=\gamma_1 K^2/(3T^2)$.



\begin{thebibliography}{20}
\bibitem{Schafer:2009dj}
T.~Sch\"afer and D.~Teaney,
``Nearly Perfect Fluidity: From Cold Atomic Gases to Hot Quark Gluon
Plasmas,''
Rept.\ Prog.\ Phys.\  {\bf 72}, 126001 (2009)
[arXiv:0904.3107 [hep-ph]].

\bibitem{Adams:2012th} 
A.~Adams, L.~D.~Carr, T.~Sch\"afer, P.~Steinberg and J.~E.~Thomas,
``Strongly Correlated Quantum Fluids: Ultracold Quantum Gases, 
Quantum Chromodynamic Plasmas, and Holographic Duality,''
New J.\ Phys.\  {\bf 14}, 115009 (2012)
[arXiv:1205.5180 [hep-th]].

\bibitem{Schaefer:2014awa} 
T.~Sch\"afer,
``Fluid Dynamics and Viscosity in Strongly Correlated Fluids,''
Ann.\ Rev.\ Nucl.\ Part.\ Sci., in press (2014)
[arXiv:1403.0653 [hep-ph]].

\bibitem{Zwerger:2016xma}
W.~Zwerger,
``Strongly Interacting Fermi Gases,''
Proceedings of the International School of Physics
``Enrico Fermi,''
Course 191,
edited by M. Inguscio, W. Ketterle and S. Stringari,
IOS Press, Amsterdam (2014)
[arXiv:1608.00457 [cond-mat.quant-gas]].

\bibitem{Schmitt:2017efp}
A.~Schmitt and P.~Shternin,
``Reaction rates and transport in neutron stars,''
Astrophys. Space Sci. Libr. \textbf{457}, 455-574 (2018)
[arXiv:1711.06520 [astro-ph.HE]].

\bibitem{Hartnoll:2021ydi}
S.~A.~Hartnoll and A.~P.~Mackenzie,
``Colloquium: Planckian dissipation in metals,''
Rev. Mod. Phys. \textbf{94}, no.4, 041002 (2022)
[arXiv:2107.07802 [cond-mat.str-el]].

\bibitem{oHara:2002}
K.~M.~O'Hara, S.~L.~Hemmer, M.~E.~Gehm, S.~R.~Granade, J.~E.~Thomas,
``Observation of a Strongly-Interacting Degenerate Fermi Gas of Atoms,''
Science {\bf 298}, 2179 (2002)
[cond-mat/0212463].

\bibitem{Kinast:2004b}
J.~Kinast, A.~Turlapov, J.~E.~Thomas,
``Breakdown of Hydrodynamics in the Radial Breathing Mode of a 
Strongly-Interacting Fermi Gas,''
Phys.\ Rev.\ A {\bf 70}, 051401(R) (2004)
[arXiv:cond-mat/0408634 [cond-mat.soft]].

\bibitem{Bartenstein:2004}
M.~Bartenstein, A.~Altmeyer, S.~Riedl, S.~Jochim, C.~Chin, 
J.~Hecker Denschlag, and R.~Grimm,
``Collective Excitations of a Degenerate Gas at the BEC-BCS Crossover,''
Phys.\ Rev.\ Lett.\ {\bf 92}, 203201 (2004) 
[cond-mat/0412712].

\bibitem{Cao:2010wa}
C.~Cao, E.~Elliott, J.~Joseph, H.~Wu, J.~Petricka, T.~Sch\"afer
and J.~E.~Thomas,
``Universal Quantum Viscosity in a Unitary Fermi Gas,''
Science {331}, 58 (2011)
[arXiv:1007.2625 [cond-mat.quant-gas]].

\bibitem{Elliott:2013}
E.~Elliott, J.~A.~Joseph, J.~E.~Thomas,
``Observation of conformal symmetry breaking and scale invariance in 
expanding Fermi gases,''
Phys.\ Rev.\ Lett.\ {\bf 112}, 040405 (2014)
[arXiv:1308.3162 [cond-mat.quant-gas]].

\bibitem{Elliott:2013b}
E.~Elliott, J.~A.~Joseph, J.~E.~Thomas,
``Anomalous minimum in the shear viscosity of a Fermi gas,''
Phys.\ Rev.\ Lett.\ {\bf 113}, 020406 (2014)
[arXiv:1311.2049 [cond-mat.quant-gas]].

\bibitem{Joseph:2015}
J.~A.~Joseph, E.~Elliott,  J.~.E.~Thomas,
``Shear Viscosity of a Unitary Fermi Gas Near the Superfluid Phase
Transition'',
Phys.\ Rev.\ Lett.\ {\bf 115}, 020401 (2015)
[arXiv:1410.4835 [cond-mat.quant-gas]].

\bibitem{Massignan:2004}
P.~Massignan, G.~M.~Bruun, H.~Smith,
``Viscous relaxation and collective oscillations in a trapped Fermi 
gas near the unitarity limit,''
Phys.\ Rev.\ A {\bf 71}, 033607 (2005) 
[cond-mat/0409660].

\bibitem{Bruun:2005}
G.~M.~Bruun, H.~Smith,
``Viscosity and thermal relaxation for a resonantly interacting 
Fermi gas,''
Phys.\ Rev.\ A {\bf 72}, 043605 (2005) 
[cond-mat/0504734].

\bibitem{Bruun:2006}
G.~M.~Bruun, H.~Smith,
``Shear viscosity and damping for a Fermi gas in the unitarity limit,''
Phys.\ Rev.\ A {\bf 75}, 043612 (2007)
[cond-mat/0612460].

\bibitem{Bruun:2007}
G.~M.~Bruun, H.~Smith,
``Frequency and damping of the Scissors Mode of a Fermi gas,''
Phys. Rev. A {\bf 76}, 045602 (2007) 
[arXiv:0709.1617].

\bibitem{Rupak:2007vp}
G.~Rupak and T.~Sch\"afer,
``Shear viscosity of a superfluid Fermi gas in the unitarity limit,''
Phys.\ Rev.\  A {\bf 76}, 053607 (2007)
[arXiv:0707.1520 [cond-mat.other]].

\bibitem{Son:2008ye} 
D.~T.~Son,
``Toward an AdS/cold atoms correspondence: A Geometric realization of 
the Schrodinger symmetry,''
Phys.\ Rev.\ D {\bf 78}, 046003 (2008)
[arXiv:0804.3972 [hep-th]].

\bibitem{Herzog:2008wg} 
C.~P.~Herzog, M.~Rangamani and S.~F.~Ross,
``Heating up Galilean holography,''
JHEP {\bf 0811}, 080 (2008)
[arXiv:0807.1099 [hep-th]].

\bibitem{Enss:2010qh} 
T.~Enss, R.~Haussmann and W.~Zwerger,
``Viscosity and scale invariance in the unitary Fermi gas,''
Annals Phys.\  {\bf 326}, 770 (2011)
[arXiv:1008.0007 [cond-mat.quant-gas]].

\bibitem{Braby:2010tk} 
M.~Braby, J.~Chao and T.~Sch\"afer,
``Viscosity spectral functions of the dilute Fermi gas in kinetic 
theory,''
New J.\ Phys.\  {\bf 13}, 035014 (2011)
[arXiv:1012.0219 [cond-mat.quant-gas]].

\bibitem{Wlazlowski:2012jb}
G.~Wlazlowski, P.~Magierski and J.~E.~Drut,
``Shear Viscosity of a Unitary Fermi Gas,
Phys. Rev. Lett. \textbf{109}, 020406 (2012)
[arXiv:1204.0270 [cond-mat.quant-gas]].

\bibitem{Hofmann:2011qs}
J.~Hofmann,
``Current response, structure factor and hydrodynamic quantities 
of a two- and three-dimensional Fermi gas from the operator product 
expansion,''
Phys.\ Rev.\ A {\bf 84}, 043603 (2011)
[arXiv:1106.6035 [cond-mat.quant-gas]].

\bibitem{Enss:2012}
T.~Enss,
``Quantum critical transport in the unitary Fermi gas,''
Phys.\ Rev.\ A {\bf 86}, 013616 (2012)
[arXiv:1204.1980 [cond-mat.quant-gas]].

\bibitem{Hofmann:2019jcj}
J.~Hofmann,
``High-temperature expansion of the viscosity in interacting quantum
gases,''
Phys.\ Rev.\ A \textbf{101}, 013620 (2020)
[arXiv:1905.05133 [cond-mat.quant-gas]].

\bibitem{Schafer:2007pr}
T.~Sch\"afer,
``The Shear Viscosity to Entropy Density Ratio of Trapped Fermions in the
Unitarity Limit,''
Phys.\ Rev.\  A {\bf 76}, 063618 (2007)
[arXiv:cond-mat/0701251].

\bibitem{Bluhm:2015bzi}
M.~Bluhm and T.~Sch{\"a}fer,
``Model-independent determination of the shear viscosity of a 
trapped unitary Fermi gas: Application to high temperature data,''
Phys. Rev. Lett. \textbf{116}, no.11, 115301 (2016)
[arXiv:1512.00862 [cond-mat.quant-gas]].

\bibitem{Bluhm:2017rnf}
M.~Bluhm, J.~Hou and T.~Sch\"afer,
``Determination of the density and temperature dependence of the shear
viscosity of a unitary Fermi gas based on hydrodynamic flow,''
Phys. Rev. Lett. \textbf{119}, no.6, 065302 (2017)
[arXiv:1704.03720 [cond-mat.quant-gas]].

\bibitem{Ku:2011}
M.~J.~H.~Ku, A.~T. ~Sommer, L.~W.~Cheuk, M.~W.~Zwierlein,
``Revealing the Superfluid Lambda Transition in the Universal
Thermodynamics of a Unitary Fermi Gas,''
Science {\bf 335}, 563 (2012)
[arXiv:1110.3309 [cond-mat.quant-gas]].

\bibitem{Kovtun:2004de}
P.~Kovtun, D.~T.~Son and A.~O.~Starinets,
``Viscosity in strongly interacting quantum field theories from 
black hole physics,''
Phys.\ Rev.\ Lett.\  {\bf 94}, 111601 (2005)
[arXiv:hep-th/0405231].

\bibitem{Hou:2021xra}
J.~Hou and T.~Sch{\"a}fer,
``Dissipative superfluid hydrodynamics for the unitary Fermi gas,''
Phys. Rev. A \textbf{104}, no.2, 023313 (2021)
[arXiv:2105.05284 [cond-mat.quant-gas]].

\bibitem{Patel:2019udb}
P.~B.~Patel, Z.~Yan, B.~Mukherjee, R.~J.~Fletcher, J.~Struck and 
M.~W.~Zwierlein,
``Universal Sound Diffusion in a Strongly Interacting Fermi Gas,''
Science \textbf{370}, no.6521, 1222-1226 (2020)
[arXiv:1909.02555 [cond-mat.quant-gas]].

\bibitem{Sommer:2011}
A.~Sommer, M.~Ku, G.~Roati, and M.~W.~Zwierlein.
``Universal spin transport in a strongly interacting Fermi gas,''
Nature {\bf 472}, 201 (2011)
[arXiv:1103.2337v1 [cond-mat.quant-gas]].

\bibitem{Baird:2019}
L.~Baird, X.~Wang, S.~Roof, J.~E.~Thomas,
``Measuring the Hydrodynamic Linear Response of a Unitary Fermi Gas,''
Phys.\ Rev.\ Lett\. {\bf 123}, 160402 (2019)
arXiv:1906.11179 [cond-mat.quant-gas]].

\bibitem{Wang:2021}
X.~Wang, X.~Li, I.~Arakelyan, J.~E.~Thomas
``Hydrodynamic Relaxation in a Strongly Interacting Fermi Gas,''
Phys\. Rev\. Lett.\, {\bf 128}, 090402 (2022)
[arXiv:2112.00549 [cond-mat.quant-gass]]

\bibitem{Yan:2022qmm}
Z.~Yan, P.~B.~Patel, B.~Mukherjee, C.~J.~Vale, R.~J.~Fletcher and 
M.~W.~Zwierlein,
``Thermography of the superfluid transition in a strongly interacting 
Fermi gas,''
Science \textbf{383}, no.6683, adg3430 (2024)
[arXiv:2212.13752 [cond-mat.quant-gas]].

\bibitem{Li:2024}
X.~Li, J.~Huang, J.~E.~Thomas, 
``Universal Density Shift Coefficients for the Thermal Conductivity 
and Shear Viscosity of a Unitary Fermi Gas,''
Phys.\ Rev.\ Res.\ {\bf 6}, L042021 (2024)
[arXiv:2402.14104 [cond-mat.quant-gas]].

\bibitem{Burnett:1935}
D. Burnett, 
``The distribution of velocities in a slightly non-uniform gas,'' 
Proc.\ Lond.\ Math.\ Soc.\ {\bf 39} 385 (1935).

\bibitem{Garcia:2008}
L.~S.~Garcia-Colin, R.~M.~Velasco, F.~J.~Uribe, 
``Beyond the Navier–Stokes equations: Burnett hydrodynamics,''
Phys.\ Rep.\ {\bf 465} 149 (2008).

\bibitem{York:2008rr} 
M.~A.~York and G.~D.~Moore,
``Second order hydrodynamic coefficients from kinetic theory,''
Phys.\ Rev.\ D {\bf 79}, 054011 (2009)
[arXiv:0811.0729 [hep-ph]].

\bibitem{Baier:2007ix} 
R.~Baier, P.~Romatschke, D.~T.~Son, A.~O.~Starinets and M.~A.~Stephanov,
``Relativistic viscous hydrodynamics, conformal invariance, and 
holography,''
JHEP {\bf 0804}, 100 (2008)
[arXiv:0712.2451 [hep-th]].

\bibitem{Chao:2011cy} 
J.~Chao and T.~Sch\"afer,
``Conformal symmetry and non-relativistic second order fluid dynamics,''
Annals Phys.\  {\bf 327}, 1852 (2012)
[arXiv:1108.4979 [hep-th]].

\bibitem{Schaefer:2014xma}
T.~Sch{\"a}fer,
``Second order fluid dynamics for the unitary Fermi gas 
from kinetic theory,''
Phys. Rev. A \textbf{90}, no.4, 043633 (2014)
[arXiv:1404.6843 [cond-mat.quant-gas]].

\bibitem{Frank:2020}
B.~Frank, W.~Zwerger, T.~Enss,
``Quantum critical thermal transport in the unitary Fermi gas,''
Phys.\ Rev.\ Res.\ {\bf 2}, 023301 (2020)
[arXiv:2003.10338 [cond-mat.quant-gas]].

\bibitem{VanHoucke:2011ux}
K.~Van Houcke, F.~Werner, E.~Kozik, N.~Prokofev, B.~Svistunov, M.~Ku, 
A.~Sommer, L.~W.~Cheuk, A.~Schirotzek and M.~W.~Zwierlein,
``Feynman diagrams versus Fermi-gas Feynman emulator,''
Nature Phys. \textbf{8}, 366 (2012)
[arXiv:1110.3747 [cond-mat.quant-gas]].

\bibitem{Nascimbene:2009}
S.~Nascimbene, N.~Navon, K.~Jiang, F.~Chevy, C~Salomon,
``Exploring the Thermodynamics of a Universal Fermi Gas,''
Nature {\bf 463}, 1057 (2010)
[arXiv:0911.0747[cond-mat.quant-gas]].

\bibitem{Dusling:2013sea}
K.~Dusling and T.~Sch{\"a}fer,
``Bulk viscosity and conformal symmetry breaking in the dilute Fermi 
gas near unitarity,''
Phys.\ Rev.\ Lett.\ \textbf{111}, no.12, 120603 (2013)
[arXiv:1305.4688 [cond-mat.quant-gas]].

\bibitem{Noronha-Hostler:2015wft}
J.~Noronha-Hostler, J.~Noronha and M.~Gyulassy,
``The unreasonable effectiveness of hydrodynamics in heavy ion collisions,''
Nucl. Phys. A \textbf{956}, 890-893 (2016)
[arXiv:1512.07135 [nucl-th]].

\bibitem{Schaefer:2014aia}
T.~Sch{\"a}fer,
``Viscosity spectral function of a scale invariant nonrelativistic 
fluid from holography,''
Phys. Rev. D \textbf{90}, no.10, 106008 (2014)
[arXiv:1408.4503 [hep-th]].

\bibitem{Son:2005tj}
D.~T.~Son,
``Vanishing bulk viscosities and conformal invariance of unitary Fermi 
gas,''
Phys.\ Rev.\ Lett.\  {\bf 98}, 020604 (2007)
[arXiv:cond-mat/0511721].

\bibitem{Chapman:1990}
S.~Chapman, T.~G.~Cowling,
``The Mathematical Theory of Non-uniform Gases,''
Cambridge University Press, 
Cambridge, UK, First edition (1939), 
Third edition (1990).

\bibitem{Braby:2010ec} 
M.~Braby, J.~Chao and T.~Sch\"afer,
``Thermal Conductivity and Sound Attenuation in Dilute Atomic Fermi 
Gases,''
Phys.\ Rev.\ A {\bf 82}, 033619 (2010)
[arXiv:1003.2601 [cond-mat.quant-gas]].

\bibitem{Bhatnagar:1954}
P.~L.~Bhatnagar, E.~P.~Gross, M.~Krook,
``A Model for Collision Processes in Gases,''
Phys.\ Rev.\ {\bf 94}, 511 (1954).

\bibitem{Loganayagam:2008is}
R.~Loganayagam,
``Entropy Current in Conformal Hydrodynamics,''
JHEP \textbf{05}, 087 (2008)
[arXiv:0801.3701 [hep-th]].

\bibitem{Jensen:2014ama}
K.~Jensen,
``Aspects of hot Galilean field theory,''
JHEP \textbf{04}, 123 (2015)
[arXiv:1411.7024 [hep-th]].

\end{thebibliography}
\end{document}